\documentclass[letter]{aa} % for the letters 
\usepackage{graphicx}
\usepackage{txfonts}
\usepackage[usenames]{color}
\usepackage{epstopdf}

\newcommand{\blue}{\textcolor{blue} }

\usepackage{threeparttable}
\usepackage{longtable}
\usepackage{lscape}
\usepackage{natbib}
\bibpunct{(}{)}{;}{a}{}{,} % to follow the A&A style

\newcommand{\teff}{ T_{{\rm eff}}}

\begin{document}
  \title{Rotational periods and evolutionary models for subgiant stars observed by 
  CoRoT\thanks{The CoRoT (Convection, Rotation and planetary Transits) space 
  mission, launched on 2006 December 27, was developed and is operated by the 
  CNES, with participation of the Science Programs of ESA, ESA's RSSD, Austria, 
  Belgium, Brazil, Germany and Spain.}
}

\subtitle{}

\author{Jos\'e-Dias do Nascimento Jr. \inst{},
              Jefferson Soares da Costa\inst{}
               \and  Matthieu Castro\inst{} 
               }
         
\offprints{J.-D. do Nascimento  Jr. }

   \institute{Depart. de F\'isica Te\'orica e Experimental, Univ. Federal do Rio 
   Grande do Norte, CEP: 59072-970 Natal, RN, Brazil \\
              \email{dias@dfte.ufrn.br}
        }

           %   \date{Received ; accepted }
\titlerunning{Rotational periods of subgiant stars}
\authorrunning {do Nascimento  et al.}
\abstract
%context heading (optional)
 { We present rotation period measurements for  subgiants  observed by CoRoT. Interpreting the modulation of  stellar light that is caused by star-spots on the time scale of the rotational period  depends on  
  knowing the fundamental   stellar parameters.} 
{Constraints on the angular momentum distribution can be extracted from the true stellar rotational 
period. By using models with an internal angular momentum distribution and comparing these with
measurements of rotation periods of subgiant stars we investigate the agreement between 
theoretical predictions and observational  results.  With this comparison we can also reduce the global 
stellar parameter space compatible with the rotational period measurements from
subgiant light curves. We can prove that an evolution assuming solid body rotation is incompatible with 
the direct measurement of the  rotational periods of subgiant stars. }
{Measuring the rotation periods relies on two different  periodogram procedures, the
Lomb-Scargle algorithm and the Plavchan  periodogram.  Angular momentum evolution models  were computed to
give us the expected  rotation periods for subgiants, which we compared with measured rotational periods.}
{We find evidence of a sinusoidal signal that is compatible in terms of both phase and amplitude with rotational 
modulation. Rotation periods were directly measured from light curves for 30 subgiant stars and  
indicate a range of 30 to 100~d for their rotational periods. }
{Our models reproduce the rotational periods obtained from CoRoT light curves. These new 
measurements of rotation periods and stellar models probe the non-rigid rotation of subgiant 
stars. \footnote{The models are  available at  http://astro.dfte.ufrn.br/prot.html~ and Table \ref{gen_table}  
and Fig. \ref{catalogue1}  are  available in  electronic form at the CDS via  
anonymous ftp to cdsarc.u-strasbg.fr  (130.79.128.5) or via 
http://cdsweb.u-strasbg.fr/cgi-bin/qcat?J/A+A/}. 
}

\maketitle
%%%%%%%%%%%%%%%%%%%%%%%%%%%%%%%%%%%%%%%%%
\section{Introduction}
%%%%%%%%%%%%%%%%%%%%%%%%%%%%%%%%%%%%%%%%%
For the first time in modern astrophysics, two space missions, Kepler and CoRoT, are providing 
accurate observations of solar-like oscillations for hundreds of main-sequence stars and thousands 
of red  giant stars  \blue{\citep{baglin06, Borucki_2009}}. CoRoT observes in the direction of the 
intersection between the equator and the Galactic plane. In each of  these fields, CoRoT has 
identified many solar-type dwarf stars \blue{\citep{baglin06}}. Another abundant class  observed  
by CoRoT are   giant stars. Intermediate between these  two groups are the subgiants,  a known rare 
class of objects. They have been accepted  as an explicit  stellar group  in 1930 
\blue{\citep{Stromberg1930}}. Until the 1950s  they were not understood in terms of stellar 
evolution theory. An interesting review of the subgiant history can be found in 
\blue{\citet{Sandage_2003}}.  Theoretically, when stars exhaust  their hydrogen in the core (turn-off), 
they leave the main sequence (MS),  pass through  the subgiant phase and become red giants.   
Studying subgiants  is interesting for several reasons:  they are particularly appropriate for 
dating purposes \blue{\citep{Thoren_2004}}, are associated with rotation periods ($P_{\mathrm {rot}}$), 
and can be  useful in stellar gyrochronology.  F-type subgiants  are  important  for studying
solar-like oscillations \blue{\citep{Barban}}. As stars evolve through the subgiant branch, their surface 
convection zone becomes deeper.   The matter that resided below the surface convection  zone 
at the MS is then exposed. From the subgiant phase to the red giant branch, the stellar radius 
increases at the same time that the convection zone  deepens. It is the first dredge-up. 
 
 	The rotation period of subgiant stars could constrain the interior angular momentum distribution and 
mixing in low-mass stars.  Subgiant $P_{\mathrm {rot}}$  measurements can teach us  about the 
rotation of low-mass stars and by implication about the solar rotation as well.  Subgiants  are more 
evolved and  slightly more  luminous than  the Sun. They are expected to rotate with $P_{\mathrm {rot}}$  
greater than the  $P_{\mathrm {rot}}^{\odot}     $.  Many of them have a convective core.  The deepening 
of the convective layers has a major influence on the chemical abundance and chromospheric activity in this phase 
(\blue{\citealt{Nascimento_2000}, 2003} and references therein). Studies show that lithium abundance 
in subgiants agrees well with dilution predictions and reflects the  well-known dilution 
process  that occurs when the  convective envelope deepens after turn-off (\blue{\citealt{Iben67}; 
\citealt{Nascimento_2000}}).  On the other hand, more massive cool subgiant stars show lithium depletion 
by up to two orders of magnitude before the start of dilution. The \blue {\citet{deMedeiros1997}; 
\citet{Lebre_1999}; \citet{Nascimento_2000} and \citet{Palacios_2003}}  results  agree with the findings by 
\blue{\citet{Balachandran_1990} and \citet{Burkhart}} about a few slightly evolved  field subgiants that originate
from the hot side of the dip and show significant  lithium depletion.  As suggested by 
\blue{\citet{Vauclair_1991}} and  \blue{\citet{Charbonnel_1992}}, the additional lithium depletion from rotationally 
\begin{figure}[t]
\includegraphics[width=9.0cm,height=9.0cm]{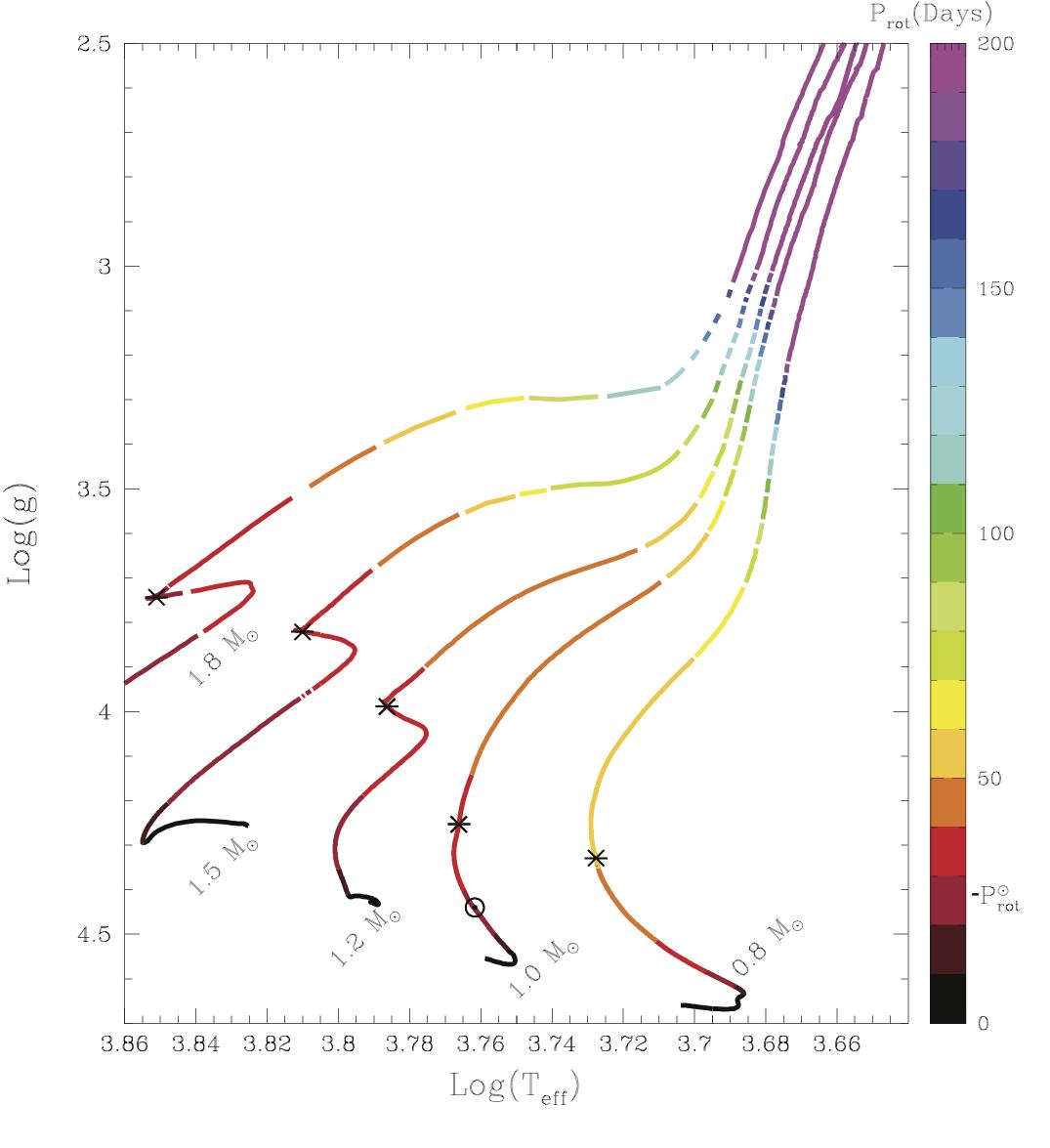}
\caption[]{ $Log(g) - \teff$  diagram.  Evolutionary models  for [Fe/H] = 0.0  and for 0.8, 1.0, 1.2, 1.5, and 1.8 solar masses. 
The color pallet denotes the evolution of rotation period for each model.  The Sun is flagged and the beginning of the subgiant 
branch (turn-off point) is marked  with an asterisk.}
\label{folding_logteff}
\end{figure}induced mixing  \blue{\citep{Charbonnel_1992, Charbonnel_1999}}  occurs  early inside these stars 
when they are on the MS, even if its signature does not appear at the stellar surface at the age of the
Hyades. This extra-mixing process is linked to rotation. If the Sun rotates as a solid body, it should have a  
$P_{\mathrm {rot}}$  greater than  90~d when it will become a subgiant. The \blue{\citet{Pinsonneaultb}} 
non standard models for subgiants imply a  $P_{\mathrm {rot}}$ of  50~d for subgiants. Clearly, a
directly measured rotation period is needed to decide in this matter. Until now, the only available  
$P_{\mathrm {rot}}$ for subgiants were inferred from chromospheric activity calibrations  \blue{\citep{Noyes_1984}} and 
are a matter of debate.  We aim to give a quantitative answer based on direct measurements of  $P_{\mathrm {rot}}$ 
for selected CoRoT subgiants.

	In this study we measured  $P_{\mathrm {rot}}$ for a sample of subgiants and present 
updated evolutionary models with an internal distribution of angular momentum to probe the expected  
$P_{\mathrm {rot}}$ for the subgiant branch.  We organize the paper as follows: we describe the observational sample 
and the $P_{\mathrm {rot}}$ determination  in Sect.\ \ref{s_obs}.   The models are presented in Sect. \ \ref{s_models}.  
The results are discussed in  Sect.\ \ref{s_res}. We summarize our  conclusions  in Sect.\ \ref{s_concl}. 
%%%%%%%%%%%%%%%%%%%%%%%%%%%%%%%%%%%%
\section{Stellar sample and time series analysis } 
\label{s_obs}
%%%%%%%%%%%%%%%%%%%%%%%%%%%%%%%%%%%%
The focus of our study  is a first analysis of the $P_{\mathrm {rot}}$ for selected subgiants observed by CoRoT. 
For the instrument description and its operation 
readers are referred to \blue{\citet{Auvergne_2009}.} 
We used  the available public data  level 2 (N2) light curves that are  ready for scientific analysis. These light curves were  delivered by the
CoRoT pipeline after  nominal corrections (\blue{\citealt{Samadi_2006}}).   In the present analysis we used 
stars classified as subgiants by \blue{\citet{Gazzano_2010}.} We chose 
only light curves with $\sim150$~d with a low contamination 
factor and high-quality  photometry  without systematic errors.
 Eventual discontinuities and long-term trends were corrected following a similar procedure as in \citet{Affer_2012}. 
%%%%%%%%%%%%%%%%%%%%%%%%%%%%%%%%%%%
\subsection{Time series analysis}
%%%%%%%%%%%%%%%%%%%%%%%%%%%%%%%%%%%

\begin{figure}[t]
\includegraphics[width=9.0cm,height=9.0cm]{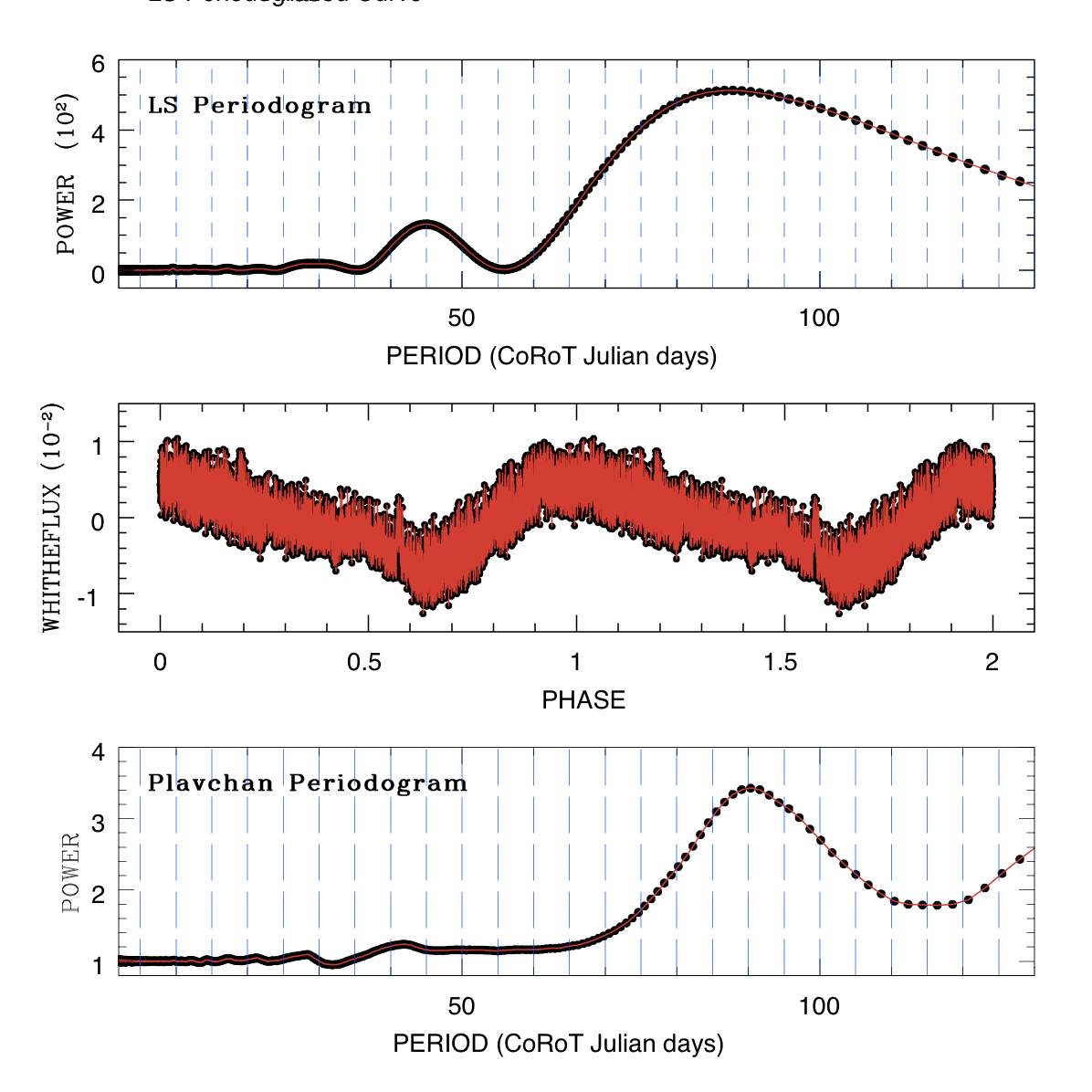}
\caption[]{CoRoT ID 100570829. In the top panel we show the LS periodogram with the
$P_{\mathrm {rot}} = 84.2$ d derived with the  LS algorithm. In the 
middle panel we show the phased curve for this $P_{\mathrm {rot}}$. In the 
bottom panel we show the Plavchan periodogram.}
\label{b1figure}
\end{figure}

	To achieve the largest possible sample of subgiants with determined $P_{\mathrm {rot}}$,  stars analyzed  
by \blue{\citet{Gazzano_2010}} that are in the  CoRoT Exo-Dat database were checked for periodic modulation. To determine
the rotational periods for our sample of subgiants,  we used a combination of two procedures, the 
Lomb-Scargle (LS) algorithm \blue{\citep{Scargle_1982}}, and the Plavchan  periodogram  \blue{\citep{Plavchan_2008}}.  
For time series in which the sampling is not uniformly distributed, the LS  periodogram analysis is particularly suitable.  
 The LS algorithm identifies  sinusoidal periodic signals in time series such 
as pulsating variable stars.  The Plavchan algorithm is a variant of the phase dispersion minimization (PDM) algorithm 
\blue{\citep{Stellingwerf_1978}} that does not use phase bins. It competently detects periodic time series shapes 
that are poorly described by the assumptions of other algorithms. This procedure is more computationally demanding than the LS analysis.

For each star, we applied the two procedures and then isolated the  most significant periods
of each method.  For the LS algorithm, we computed the normalized power as a function of periods and then searched for peaks in the 
 function.  To decide  whether there is a significant signal from a certain period in the power spectrum,  the power at that period was linked 
to the false-alarm probability (FAP). This is the probability that a peak with a power z equal to, or higher than, the highest peak observed 
in the periodogram would appear anywhere in the considered frequency range in the presence of pure noise.
We derived the FAP  for all detected periods and only those  with an FAP $<$ 0.000001 were considered as 
significant peaks.  The FAP is given by  $FAP = 1 - [1 - exp(-z)]^{N_i}$, where  $N_i  = -6.362 +1.193N + 0.00098N^{2}$  
 is the number of independent frequencies, $N$ is the number of data points and $z$  the height of the peak
 (e.g., \blue{\citealt{Horne}}). For clumped data, \blue{\citet{Scholz_2004}}  used $N_i$ = N / 2 for their first  estimation of an FAP.  
These  two different approaches do not result in a significant change of the FAP.
The light curve coverage allows us to detect periods longer than 2~d 
and shorter than 100~d  with a relevant FAP  $<$ 0.000001.  The uncertainties in   $P_{\mathrm {rot}}$  are determined by the frequency 
resolution in the power spectrum and the sampling error.  The error of our 
measurement is defined by the  probable error, which in turn is defined as 0.2865$\cdot$FWHM (full-width at half-maximum of the peak),
assuming  a Gaussian statistics around the LS peak. 
The most significant periods are compared with models presented in Sect. \ref{s_models}.  
Our final work sample is composed of 30 subgiant stars. The derived periods and their respective errors are presented in Table \ref{gen_table}.
In Fig.~\ref{b1figure} we show the Lomb-Scargle periodogram (top), the  phased  curve (middle), and the 
Plavchan periodogram (bottom) for CoRoT ID 100570829.  This is a representative subgiant star classified as a K0IV
in the Exo-Dat (\blue{\citealt{Deleuil_2006}, 2009; \citealt{Meunier_2007}})  and \blue{ \citet{Gazzano_2010}}. The 
data we used  were obtained during the CoRoT first long run (LRc01 and LRa01).  The rotation periods for this target were derived from
Lomb-Scargle and Plavchan periodograms analysis of the light curve, giving $P_{\mathrm {rot}} =  84.2~\pm $  15.8 d with
a FWHM of 55 d. The $P_{\mathrm {rot}}$  uncertainty comes mainly from the time series limitation.
\begin{figure}[t]
\includegraphics[width=9.0cm,height=9.0cm]{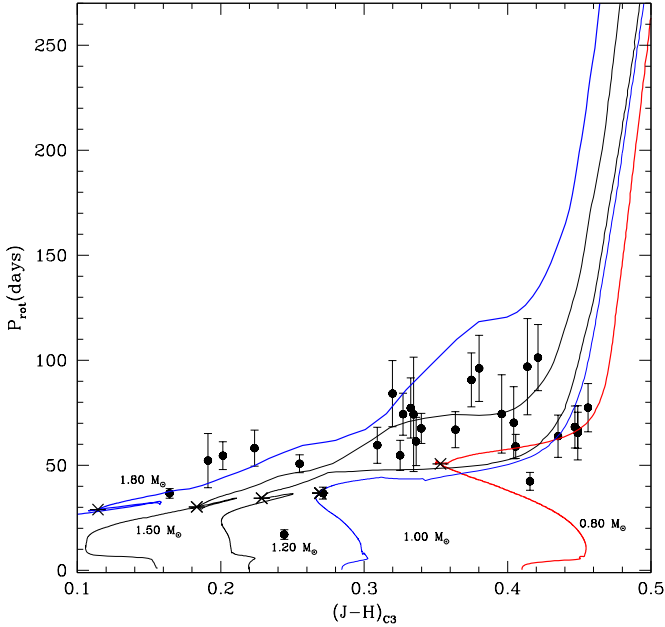}
\caption[]{Rotation period evolution from the main sequence to the giant branch as a function of (J -- H)$_{C3}$ index.  The models are shown for 
[Fe/H] = 0 and different stellar masses computed with differential rotation. 
Filled circles stand for rotational periods determined  as described in the  Sect.\ \ref{s_obs}.
The beginning of the subgiant branch (turn-off point) is flagged  with an asterisk.}
\label{Prot_JH}
\end{figure}

\begin{figure}[t]
\includegraphics[width=9.0cm,height=9.0cm]{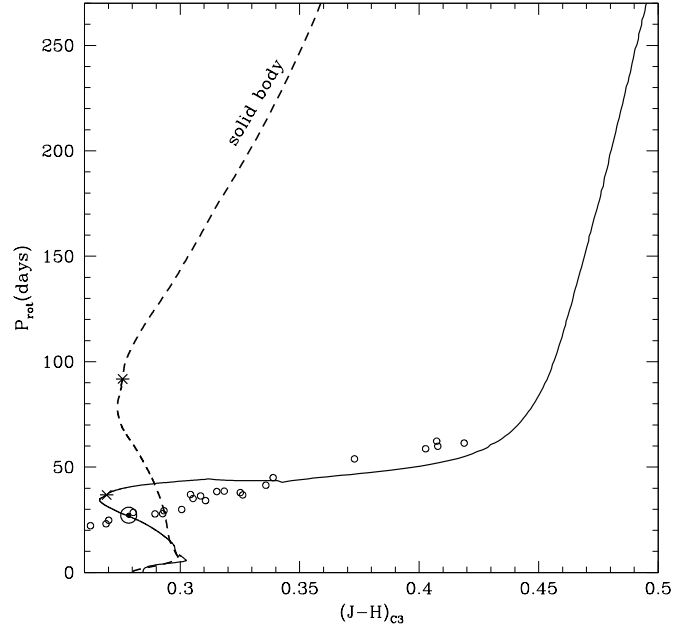}
\caption[]{Rotation period evolution from the main sequence to the giant branch as a function of the (J -- H)$_{C3}$ index.  
The models are shown for [Fe/H] = 0 and 1 M$_{\odot}$. The dashed line represents the rotation period evolution 
for a solid-body model. The  solid line represents the computed model with differential rotation. 
Open circles represent  subgiants with  masses of about one solar mass and 
rotational periods determined from the rotation-activity relation by \blue{\citet{Lovis_2011}}. Asterisk as in Fig~\ref{Prot_JH}.}
\label{1solarmodel}
\end{figure}

%%%%%%%%%%%%%%%%%%%%%%%
\section{Evolutionary models}
%%%%%%%%%%%%%%%%%%%%%%%%
\label{s_models}
Our models\footnote{available at  http://astro.dfte.ufrn.br/prot.html} were computed with the Toulouse-Geneva stellar evolution code  \blue{\citep{huibonhoa_2008}}. More details of the physics used in the models 
can be found in \blue{\citet{Richard_1996,Richard_2004}}, \blue{\citet{huibonhoa_2008}}, and \blue{\citet{Nascimento_2009}} as well as in Appendix~\ref{apendixb} . The initial composition follows the \blue{\citet{Grevesse_93}} mixture. The convection was treated according to the \blue{\citet{bohmvitense}} formalism of the mixing length theory with a mixing length parameter $\alpha = l/H_{\mathrm p}$, where $l$ is the mixing length and $H_{\mathrm p}$ the pressure height scale. The rotation-induced mixing due to meridional circulation and the transport of angular momentum due to rotationally induced instabilities are computed as described by  \blue{\citet{Zahn_1992}}  and \blue{\citet{Talon_1997}} and takes into account differential rotation. The angular momentum distribution at a given time is a function of its previous history. As underlined by \blue{\citet{Pinsonneaultb}}, the initial conditions are critical for rotating models. The angular momentum loss  from the disk-locking  process is linked  with stellar magnetic fields and remains poorly understood. The evolution of the angular momentum is computed with the \blue{\citet{Kawaler_1988}} prescription as in equation \ref{equ:kawaler}. Our solar model is calibrated to match the observed solar effective temperature, luminosity, and rotation at the solar age. The calibration is based on the \blue{\citet{Richard_1996}} prescription. For a 1.0 $M_{\odot}$ star, we adjusted the mixing-length 
parameter $\alpha$ and the initial helium abundance $Y_{\mathrm{ini}}$ to reproduce the observed solar luminosity, and the radius at the solar age: 
$L_{\odot} = 3.8515 \pm 0.0055 \times 10^{33} \ \mathrm{erg \cdot  s^{-1}}$, $R_{\odot} = 6.95749 \pm 0.00241 \times 10^{10} 
\ \mathrm{cm}$, and $t_{\odot} = 4.57 \pm 0.02 \ \mathrm{Gyr}$ \blue{\citep{Richard_2004}}. For the best-fit solar model, we obtained 
$L = 3.8514 \times 10^{33} \ \mathrm{erg \cdot s^{-1}}$ and $R = 6.95750 \times 10^{10} \ \mathrm{cm}$ at an age $t = 4.576 \ \mathrm{Gyr}$, with ${Y_{\mathrm{ini}} = 0.277}$ and ${\alpha = 1.72}$. The  free parameters of  the  rotation-induced  mixing determine  the efficiency of the  turbulent motions and are adjusted to produce a mixing that satisfies both the helium gradient below the surface convective zone, which improves the agreement between the model and seismic sound speed profiles, and the absence of beryllium destruction  (see Appendix~\ref{apendixb} for details. We used the initial angular momentum inferred by \blue{\citet{Pinsonneaultb}} for the Sun $J_0 = 1.63 \times 10^{50} \mathrm{~g \cdot cm^2 \cdot s^{-1}}$ and calibrated the angular momentum loss by requiring that the solar model has the solar rotation rate at the solar age. We obtained the solar surface rotation velocity 
$v = 2.2 \mathrm{~km \cdot s^{-1}}$ at the solar age.  The models of  0.8, 1.0, 1.2, 1.5, and 1.8  solar masses (Fig.~\ref{folding_logteff}) were computed from the zero-age main 
sequence (ZAMS) to the top of the red giant branch (RGB) with the same calibration  values as  for the solar model. 
In these diagrams the asterisk indicates the evolutionary region where the subgiant branch starts, this point 
corresponds to the age of hydrogen exhaustion  in stellar central regions.  From \blue{\citet{Alonso_1999}} we calibrated  
$\teff$ as function of (J -- H) color index. For the selected objects we obtained (J -- H)  2MASS photometry  from  CoRoT 
Exo-Dat and applied the pseudo-colors $C_3$ reddening correction as described by \blue{\citet{Catelan2011}} to obtain the  
 reddening-free index (J -- H)$_{C3}$.

%%%%%%%%%%%%%%%%%%%%%%%%%%%%%%%%%%%%%%%%%%%%%%%%%%
\section{Results}
\label{s_res}
%%%%%%%%%%%%%%%%%%%%%%%%%%%%%%%%%%%%%%%%%%%%%%%%%%
The $P_{\mathrm {rot}}$ measured from light curves for subgiants are shown in Fig.~\ref{Prot_JH}. 
These $P_{\mathrm {rot}}$ measurements  indicate the range of 30 to 100~d for subgiants in
agreement with expected  $P_{\mathrm {rot}}$ from models. The $P_{\mathrm {rot}}$  for subgiants 
with $M < 1.80~M_{\odot}$  increases slightly until the bottom of the RGB. Fig.~\ref{1solarmodel}  compares 
the expected $P_{\mathrm {rot}}$ from a 1.00 $M_{\odot}$ model rotating as a solid body with  a
1.00 $M_{\odot}$  rotating differentially.   Open circles represent  $P_{\mathrm {rot}}$  for  subgiants determined
by  \blue{\citet{Lovis_2011}}.  Our models follow the same prescription for the angular momentum loss as 
 \blue{\citet{Pinsonneaultb}} and an updated physics (opacities, equation of state, meridional circulation, and shear instabilities).
  The metallicity effect on the evolution of the rotational period for  models with   $[Fe/H] = \pm 0.3~dex $ 
is lower  than the LS intrinsical error. 
 On the subgiant branch, the separations of solid-body rotation models and models with internal angular 
momentum distribution are satisfactorily distinguished by the  measured rotation periods from 
CoRoT light curves.  Thus, we emphasize that this range of rotation periods 
for subgiants reinforces the scene of a strong radial differential rotation in depth at the main sequence
or a fast core rotating in the subgiant branch.  The $P_{\mathrm {rot}}$  in the subgiant phase is driven by the deepening of the 
convective zone, which  extracts angular momentum from  the radial differential rotation reservoir. 
From the  angular momentum conservation, this extraction compensates for the increase of the momentum of inertia due to 
the stellar radius enhancement. It causes  
the difference  between the two models shown in Fig.~\ref{1solarmodel}.
Another  interesting fact is that subgiants present a chromospheric activity lower than main-sequence 
stars with the same mass \blue{\citep{Lovis_2011}}, even if its convective zone is deeper than their progenitor. 
This scenario contrasts with  the suggestion of the magnetic breaking as the root cause of the 
low rotation of subgiants \blue{\citep{Gray_1985}}.
 %%%%%%%%%%%%%%%%%%%%%%%%%%%%%%%%%

\section{Conclusion}
\label{s_concl}
We have reported $P_{\mathrm {rot}}$  for 30  subgiants observed by CoRoT. 
These $P_{\mathrm {rot}}$  combined with evolutionary  models 
helped us to more tightly constrain the angular momentum evolution for evolved stars, which is inaccessible 
to direct observations. Our  $P_{\mathrm {rot}}$  agree well with the range of periods  determined 
from the activity modulation studies of subgiants. We showed that subgiants  present  $P_{\mathrm {rot}}$  
ranging from  30 to 100~d in the mass range from 0.8 to 1.8 solar masses.
This work presents a first step in addressing the study of the  $P_{\mathrm {rot}}$ of the
subgiants.  Our models agree with rotational period measurements for subgiant stars.
The rotation period range  for subgiants reinforces the scene of a differential rotation in depth or a fast-core rotating.  
This result also agrees  with the findings by \blue{\citet{Mosser_2012}} who observed that rotational 
splittings and  core rotation significantly slows down during the RGB.

\begin{acknowledgements}
We would like to  thank the Exo-Dat staff. This work has been supported by 
 the CNPq Brazilian  Agency CNPq/PQ fellowship. JDNJr dedicates this work 
 to his children, Malu  and L\'eo.  An anonymous referee is acknowledged for helpful suggestions.

  \end{acknowledgements}

%%%%%%%%%%%%%%%%%%%%%%%%%%%%%%%%%%%%%%%%%%%%%%%%%%%%%%%%%%%%%%
\bibliographystyle{aa} % style aa.bst
\bibliography{bibi} % your references Yourfile.bib

\newpage 

\appendix

 \section{Description of the physics adopted for the transport of internal angular momentum}
\label{apendixb}
In addition to the input described in Sect.\ \ref{s_models}., we provide here more details on the physics adopted for the transport of internal angular momentum for the present modeling.  We used the  same approach as \blue{\citet{Pinsonneaultb}} after updating the treatment of the instabilities relevant to the transport of angular momentum according to \blue{\citet{Zahn_1992}} and \blue{\citet{Talon_1997}}.\\

 {\it Initial conditions}: The rotational properties are strongly influenced by the pre-main-sequence phase. \blue{\citet{Kawaler_1987}} determined the initial angular momentum for the Sun as $J_0 = 1.63 \times 10^{50} \ \mathrm{g \cdot cm^{2} \cdot s^{-1}}$.\\
 
 {\it Angular momentum loss}:  \blue{\citet{Kawaler_1988}} described the angular momentum loss for stars with an outer convective envelope as
\begin{equation}
 \frac{dJ}{dt} = -K \Omega^{1+4N/3} \left(\frac{R}{R_{\odot}}\right)^{2-N} \left(\frac{M}{M_{\odot}}\right)^{-N/3}, %\left(\frac{\dot{M}}{10^{-14}}\right)^{1-2N/3}
\label{equ:kawaler}
\end{equation}
with $\Omega$ the angular velocity and $K$ a constant that combines scale factors for the wind velocity and magnetic field strength. This is adjusted to give the solar surface rotation velocity at the solar age. $N$ denotes the wind index and is a measure of the magnetic geometry. $N=3/2$ for the Sun. The  mass loss dependence rate is very weak, and we assumed the rate $\dot{M}$ to be $10^{-14} \ M_{\odot} \cdot \mathrm{yr^{-1}}$.\\

{\it Transport of matter and angular momentum}: The redistribution of matter and angular momentum is carried out by dynamical instabilities (convection and dynamical shear mainly) that occur on a time scale much shorter than the evolutionary time scale, and also by secular instabilities (Eddington circulation and secular shear) with a similar or longer time scale. The Eddington meridional circulation \blue{\citep{Eddington26}}, is a large-scale mass motion due to thermic gradients caused by rotation.  The vertical velocity $U_{\mathrm r}$ of this circulation is related to the divergence of the radiation flux \citep{Eddington25,Sweet50,Zahn_1987}. In a uniformly rotating star, $U_{\mathrm r}$ has the following analytical form:
\begin{equation}
 U_r = \frac{8}{3} \frac{\Omega^2 r^5}{G^2} \frac{L}{M^3} \frac{\nabla_{\mathrm{ad}}}{\nabla_{\mathrm{ad}} - \nabla_{\mathrm{rad}}} \left( 1 - \frac{\Omega^2}{2 \pi G \rho} \right) P_2 (\cos \theta),
 \end{equation}
 where $\Omega$ is the angular velocity, $r$ the mean radius, and $\rho$ the density of the considered equipotential, $L$ and $M$ are the luminosity and the mass at this location, $G$ is the gravitational constant, $\nabla_{\mathrm{ad}}$ and $\nabla_{\mathrm{rad}}$ are the adiabatic and radiative gradients; $P_2  (\cos \theta)$ is the second-order Legendre polynomial in which $\theta$ is the colatitude. This flow advects angular momentum and thereby induces differential rotation. \\

The rotation state is then a result from the balance between meridional advection and turbulent stresses. Shear instabilities ensure that the angular velocity is constant in equipotential surfaces. The turbulent viscosity is assumed to be anisotropic and dominant in the horizontal over the vertical direction. Horizontal turbulent motions work against the advection of chemicals by the meridian flow, which homogenizes horizontal layers. The vertical transport of matter is accordingly treated as a diffusion process:
\begin{equation}
 \rho \frac{\partial \bar{c}}{\partial t} = \frac{1}{r^{2}} \frac{\partial}{\partial r} \left( r^{2} \rho D_{turb} \frac{\partial \bar{c}}{\partial r} \right) ,
\end{equation} 
with $\rho$ being the density, $r$ the radial coordinate, $\bar{c}$ the mean concentration diffusing vertically, and \\

$D_{turb} = D_{v} + \frac{[r U_{r} (r)]^{2}}{30 D_{h}}$ \\

the turbulent diffusion coefficient expressed from the vertical and horizontal diffusion coefficients, valid when $D_{h} \gg D_{v}$. $U_{r} (r)$ is the amplitude of the vertical component of the circulation velocity. If we assume that the meridional velocity and the turbulent diffusion coefficients are correlated \blue{\citep{Zahn_1992}}, we 
have \\

$D_{turb} = \alpha_{turb} r |U_r|$, with $\alpha_{turb} = C_v + \frac{1}{30 C_h}$. \\

 The free parameters 
$\alpha_{turb}$ and $C_h$ are adjusted in our models to reproduce the solar proprieties affected by rotation-induced mixing. The values found in the 
calibration presented in Sect.\ \ref{s_models} are $\alpha_{turb} = 1$ and $C_h = 9000$.\\

The transport of angular momentum is governed by an advection/diffusion equation:
\begin{equation}
 \frac{\partial}{\partial t} \left[ \rho r^{2} \Omega \right] = \frac{1}{5r^{2}} \frac{\partial}{\partial r} \left[ \rho r^{4} \Omega U_{r} \right] + \frac{1}{r^{2}} \frac{\partial}{\partial r} \left[ \rho \nu_v  r^{4} \frac{\partial \Omega}{\partial r} \right],
\end{equation} 
with $\Omega$ the angular velocity and $\nu_v$ the vertical turbulent viscosity. We use the prescription given by \blue{\citet{Talon_1997}}:
\begin{equation}
 \nu_v = D_v = \frac{8 Ri_c}{5} \frac{(r d \Omega/dr)^{2}}{N_T^{2}/(\kappa+D_h)},
\end{equation}
taking into account the homogenizing effect of the horizontal diffusion ($D_h$) on the restoring force caused by the gradient of molecular weight.~$N_T^{2} = \frac{g \delta}{H_P} (\nabla_{ad} - \nabla)$ is the Brunt-V\"ais\"al\"a frequency, $\kappa$ the radiative diffusivity, and $Ri_c \sim 1/4$ is the critical Richardson number \blue{\citep[see][]{Talon97}}. The horizontal shear is sustained by the advection of momentum:
\begin{equation}
 D_h = \frac{rU_{r}}{C_h} \left[ \frac{1}{3} \frac{d \ln \rho r^{2} U_{r}}{d \ln r} - \frac{1}{2} \frac{d \ln \rho r^{2} \Omega}{d \ln r} \right].
\end{equation}

\newpage

\onecolumn
\longtab{1}{
\begin{longtable}{ c c c c c c c c c c  }
\caption[table]{The subgiant sample} 
\\
\hline 
CoRoT ID & RA & DEC & $(J-H)_{C3}$ & $P_{\mathrm{rot}}$ (LS) &   error & FAP & FWHM & $P_{\mathrm{rot}}$ (Plavchan) &   \\
 & (deg) & (deg) & (mag) & (days) &  (days) & & (days) & (days) &  \\
\hline 
\endhead
\hline 
\endlastfoot
\\

%  ID                                   RA                                    DEC                          C3                         Pls                         Erro                   FAP          FHWD             PDM               
100519170        &        290.743009        &        1.406339        &        0.33446        &        74.3       &                27.2        &    $<$ 10$^{-6}$   &       95        &        73.2               \\
100543054        &        290.778772        &        1.326228        &        0.43519        &        63.9       &                10.0        &     $<$ 10$^{-6}$   &      35        &        65.6               \\
100570829        &        290.820823        &        1.19351          &        0.31981        &        84.2       &                15.8        &     $<$ 10$^{-6}$   &     55        &        87.5               \\
100603128        &        290.869237        &        1.430996        &        0.44692        &        72          &                10.0        &     $<$ 10$^{-6}$   &      35        &        62.9               \\
100686488        &        290.994464        &        0.948138        &        0.44898        &        66.7       &                8.6          &     $<$ 10$^{-6}$   &    30        &        69.0               \\
100698726        &        291.015259        &        1.174825        &        0.24425        &        17.1       &                2.3          &     $<$ 10$^{-6}$   &   8        &        16.9               \\
100722142        &        291.055588        &        1.225582        &        0.32506        &        54.8       &                7.2          &     $<$ 10$^{-6}$   &   25        &        52.3               \\
100723404        &        291.058201        &        1.394501        &        0.22351        &        58.2       &                8.6          &     $<$ 10$^{-6}$   &    30        &        60.8               \\
100726847        &        291.066425        &        0.671149        &        0.32713        &        74.4       &                10.0        &    $<$ 10$^{-6}$   &     35        &        78.1               \\
100736747        &        291.081046        &        1.24076          &        0.44892        &        65.5       &                12.9        &    $<$ 10$^{-6}$   &    45        &        70.3               \\
100754501        &        291.104633        &        0.868713        &        0.16429        &        36.7       &                2.3          &     $<$ 10$^{-6}$   &   8        &        35.8               \\
100796424        &        291.160582        &        0.883204        &        0.4138          &        97           &                22.9        &    $<$ 10$^{-6}$   &    80        &        100.3               \\
100799876        &        291.165061        &        0.696372        &        0.40434        &        70.3       &                17.2        &    $<$ 10$^{-6}$   &    60        &        64.0               \\
100820430        &        291.192124        &        0.626198        &        0.41559        &        42.4       &                4.3           &    $<$ 10$^{-6}$   &  15        &        41.9               \\
100840079        &        291.217916        &        0.846178        &        0.40562        &        59          &                5.7           &    $<$ 10$^{-6}$   &  20        &        61.6               \\
100894594        &        291.289809        &        1.475675        &        0.1911          &        52.3       &                12.9         &    $<$ 10$^{-6}$   &    45        &        57.6               \\
100914011        &        291.315583        &        1.610448        &        0.42116        &        101.3    &                15.8         &   $<$ 10$^{-6}$   &     55        &        97.0               \\
101119921        &        291.641478        &        0.665546        &        0.27134        &        36.8       &                2.9          &    $<$ 10$^{-6}$   &    10        &        37.7               \\
101139463        &        291.672001        &        0.561325        &        0.3362          &        61.4       &                11.5        &   $<$ 10$^{-6}$   &     40        &        57.5               \\
101195094        &        291.759369        &        0.543763        &        0.33248        &        77.3       &                14.3        &   $<$ 10$^{-6}$   &     50        &        65.7               \\
101237986        &        291.826509        &        0.524885        &        0.33984        &        67.6       &                7.2          &   $<$ 10$^{-6}$   &     25        &        67.2               \\
101297209        &        291.917861        &        0.698688        &        0.30915        &        59.6       &                8.6           &    $<$ 10$^{-6}$   &    30        &        63.0               \\
101541502        &        292.358882        &        -0.014249       &        0.255            &        50.8        &                4.3          &   $<$ 10$^{-6}$   &     15        &        50.2               \\
101642233        &        292.544083        &        0.042465        &        0.37479        &        90.7       &                12.9         &   $<$ 10$^{-6}$   &     45        &        90.6               \\
102627000        &        100.520209        &        -1.04878         &        0.56943        &        52.8       &                 5.7           &     $<$ 10$^{-6}$   &   20        &        57.3               \\
102636100        &        100.575778        &        -1.250857       &        0.39593        &        74          &                18.6         &     $<$ 10$^{-6}$   &   65        &        99.4               \\
102645577        &        100.626996        &        -1.002883       &        0.38023        &        96.2       &                15.8        &   $<$ 10$^{-6}$   &     55        &        90.8               \\
102690215        &        100.860987        &        -1.183079       &        0.20153       &        54.6        &                 6.6           &      $<$ 10$^{-6}$   &  23        &        57.5               \\
102736038        &        101.115724        &        -0.946626       &        0.36367        &        67          &                 8.6            &    $<$ 10$^{-6}$   &    30        &        60.4               \\
110603474        &        291.002295        &        0.748182         &        0.45604        &        77.5      &                11.5             &     $<$ 10$^{-6}$   &   40        &        75.6               \\

\vspace{0.3pt}
\label{gen_table}

\end{longtable}                                                                           
}

\begin{figure*}[]

\includegraphics[angle=0,width=4.5cm]{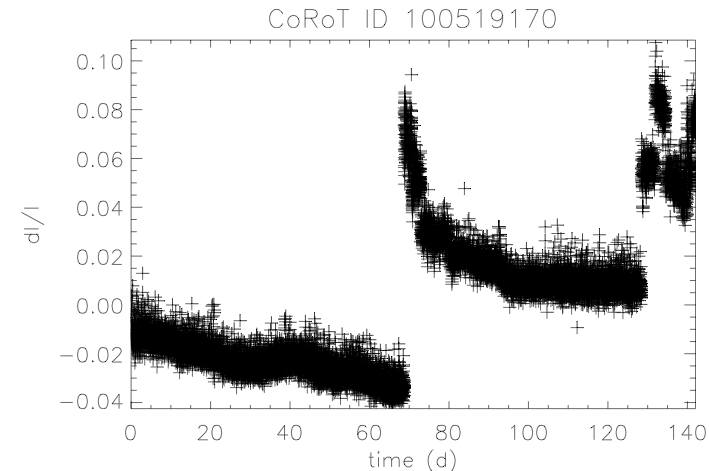}
\includegraphics[angle=0,width=4.5cm]{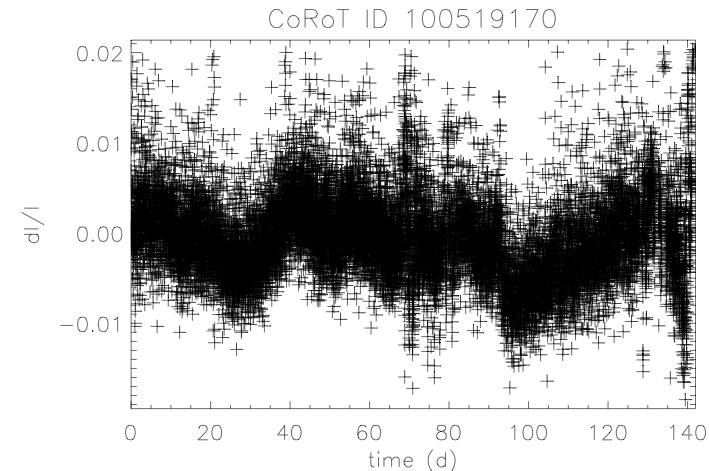}
\includegraphics[angle=0,width=4.5cm]{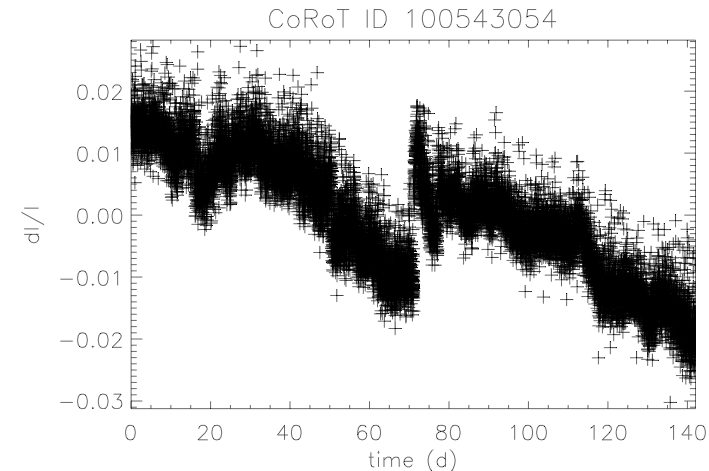}
\includegraphics[angle=0,width=4.5cm]{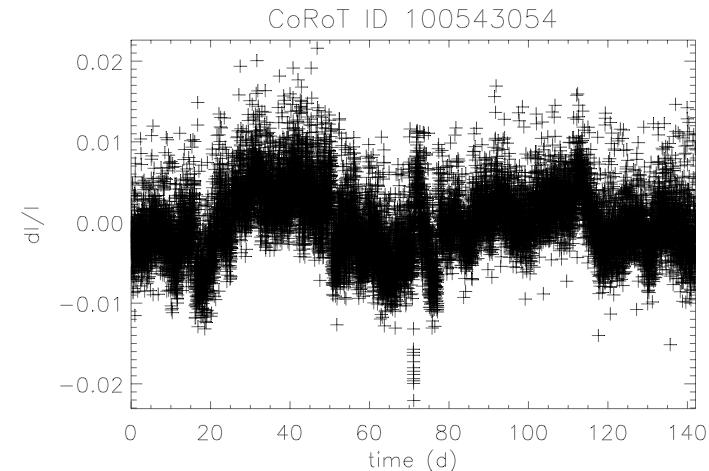}
\includegraphics[angle=0,width=4.5cm]{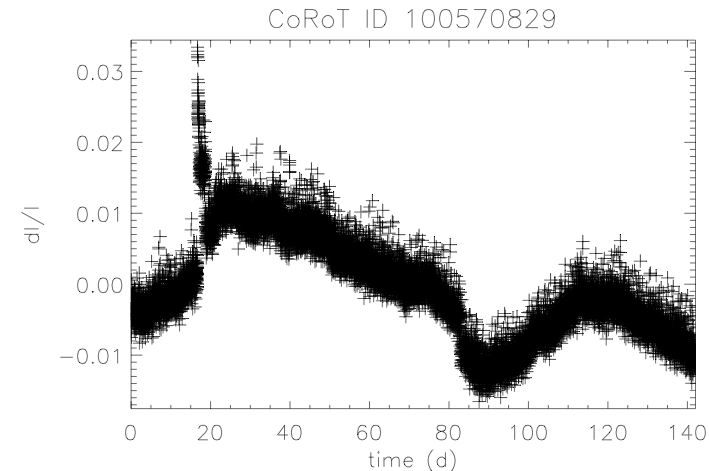}
\includegraphics[angle=0,width=4.5cm]{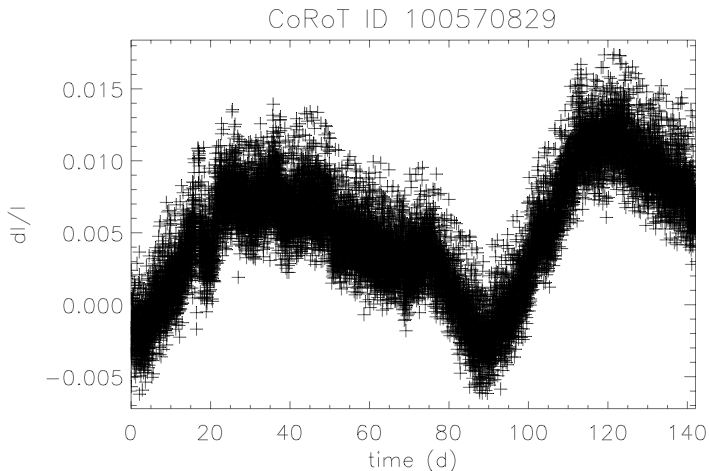}
\includegraphics[angle=0,width=4.5cm]{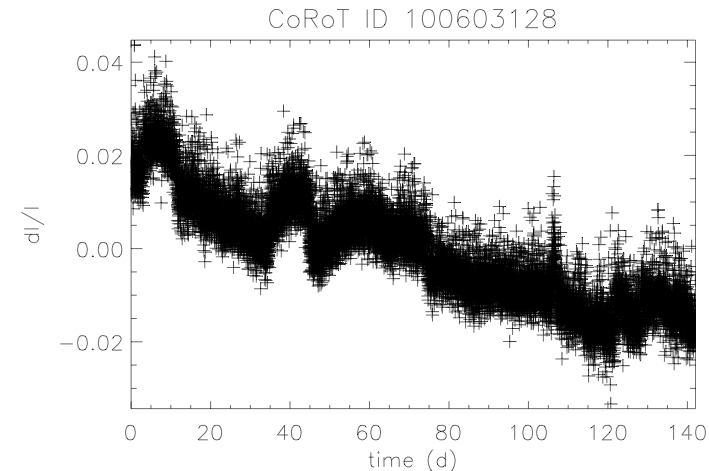}
\includegraphics[angle=0,width=4.5cm]{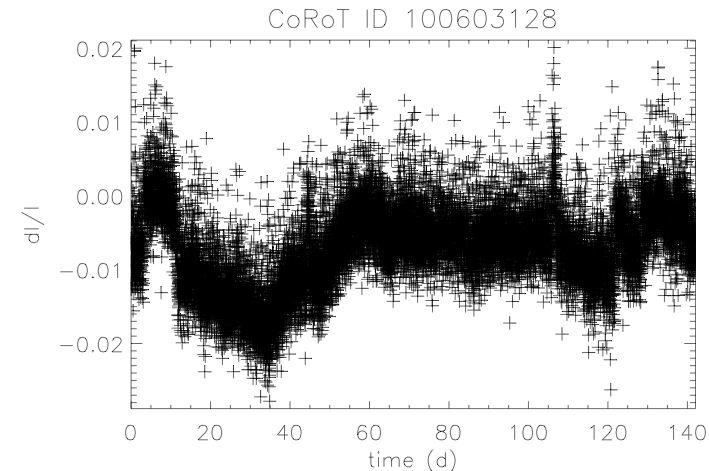}
\includegraphics[angle=0,width=4.5cm]{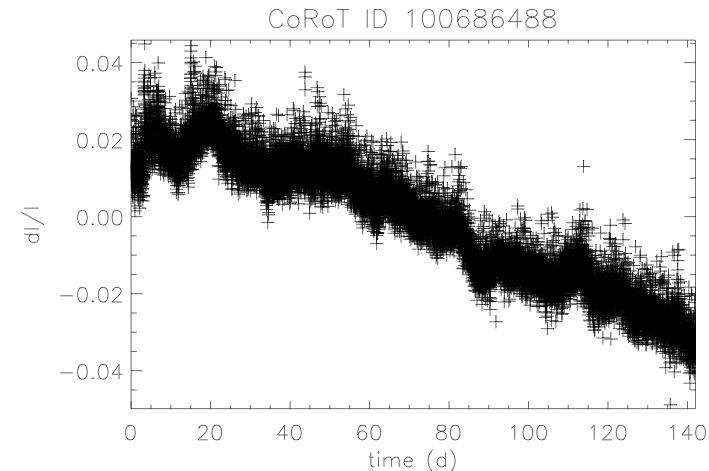}
\includegraphics[angle=0,width=4.5cm]{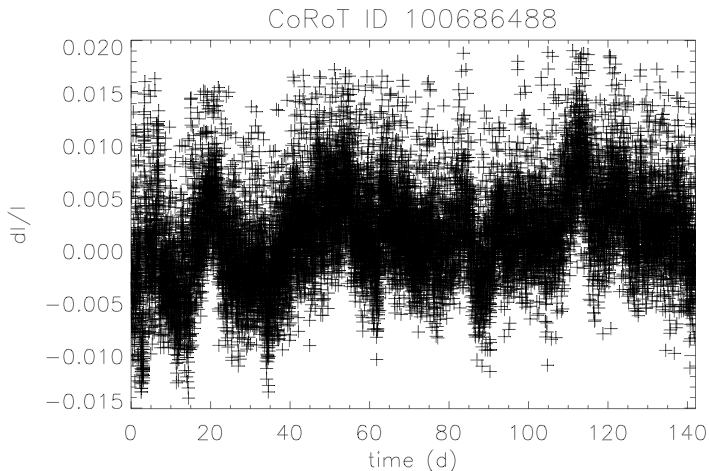}
\includegraphics[angle=0,width=4.5cm]{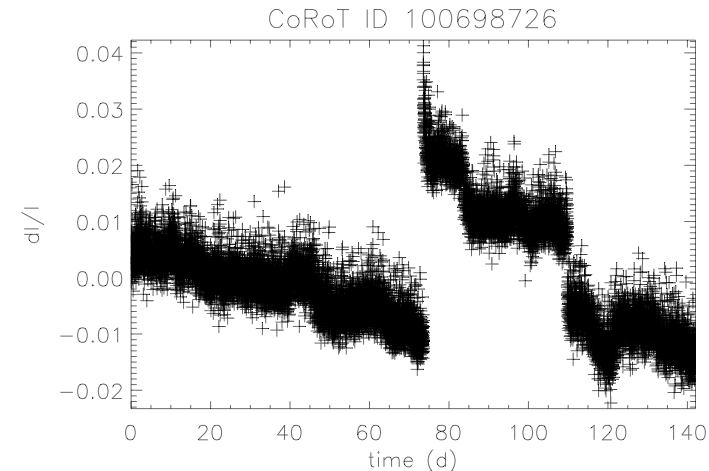}
\includegraphics[angle=0,width=4.5cm]{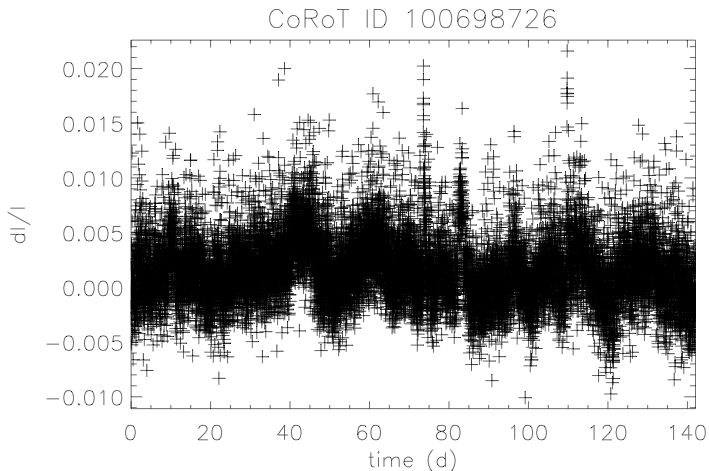}
\includegraphics[angle=0,width=4.5cm]{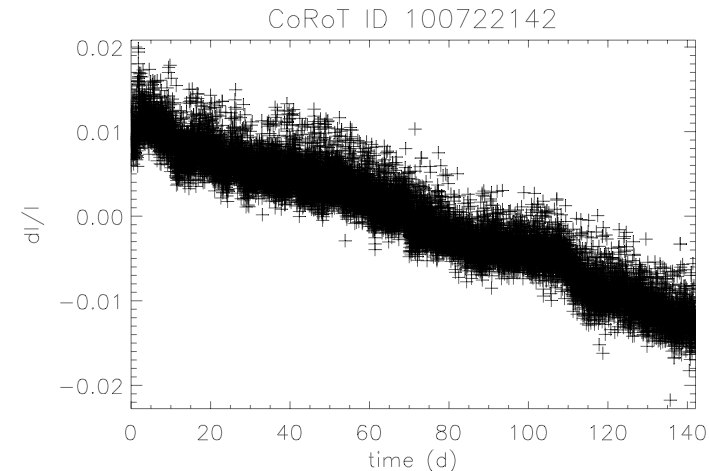}
\includegraphics[angle=0,width=4.5cm]{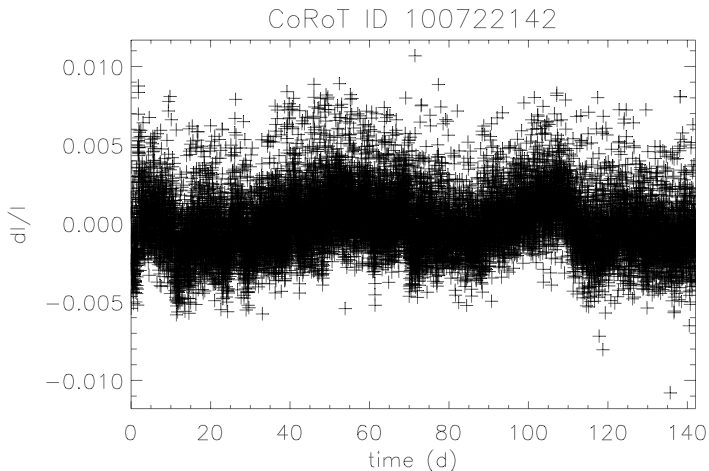}
\includegraphics[angle=0,width=4.5cm]{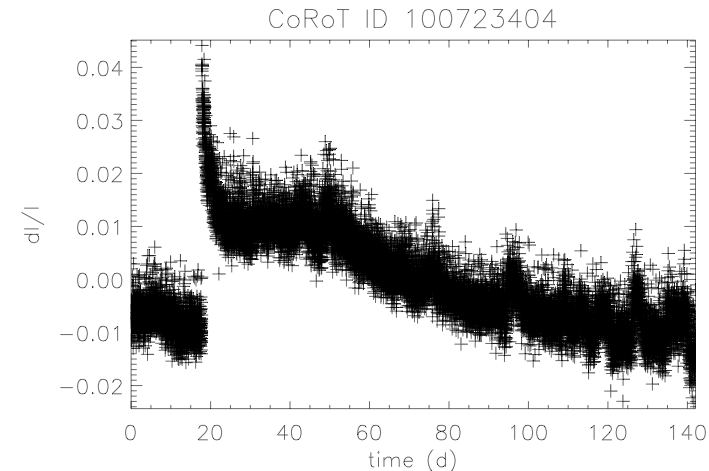}
\includegraphics[angle=0,width=4.5cm]{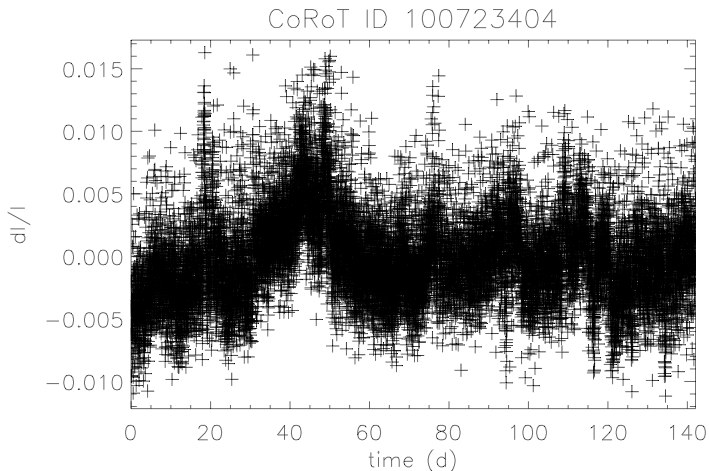}
\includegraphics[angle=0,width=4.5cm]{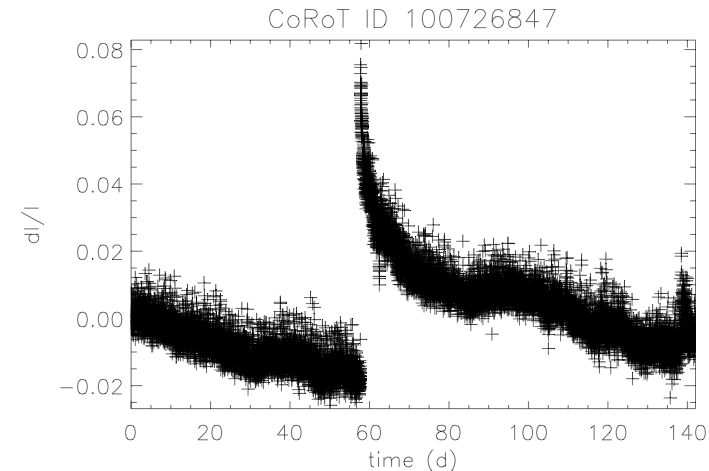}
\includegraphics[angle=0,width=4.5cm]{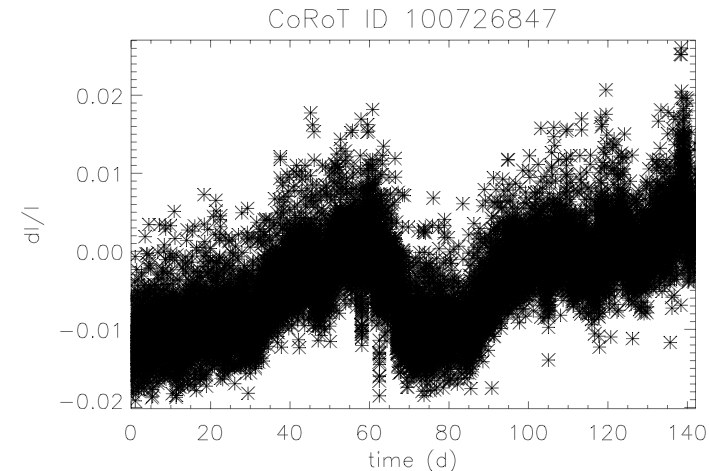}
\includegraphics[angle=0,width=4.5cm]{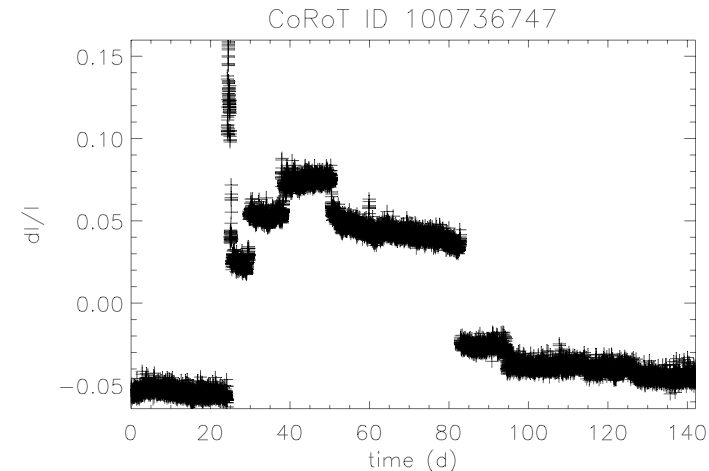}
\includegraphics[angle=0,width=4.5cm]{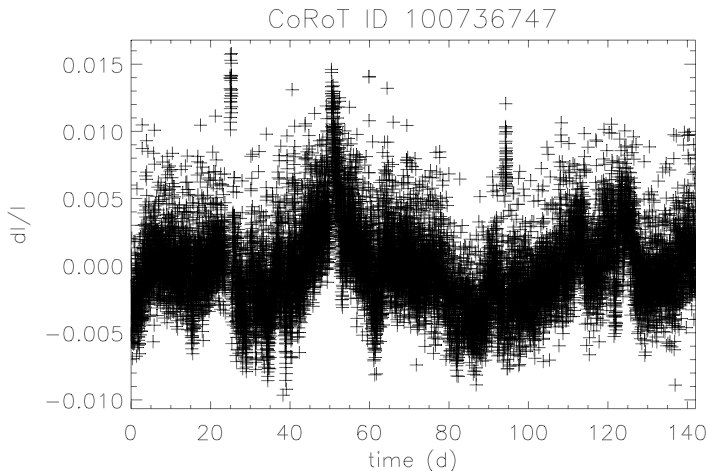}
\includegraphics[angle=0,width=4.5cm]{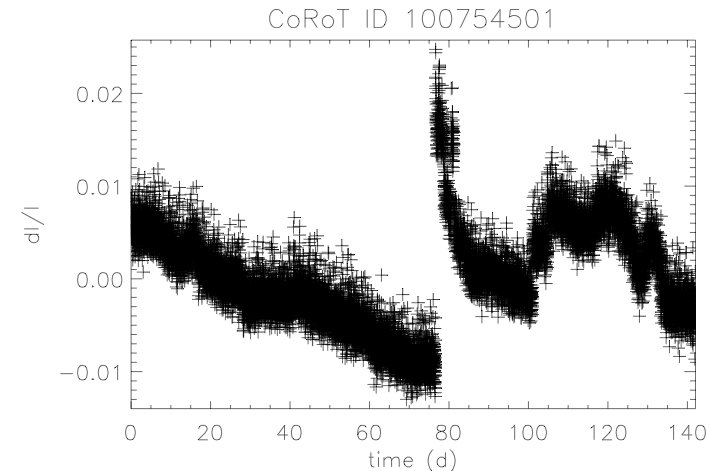}
\includegraphics[angle=0,width=4.5cm]{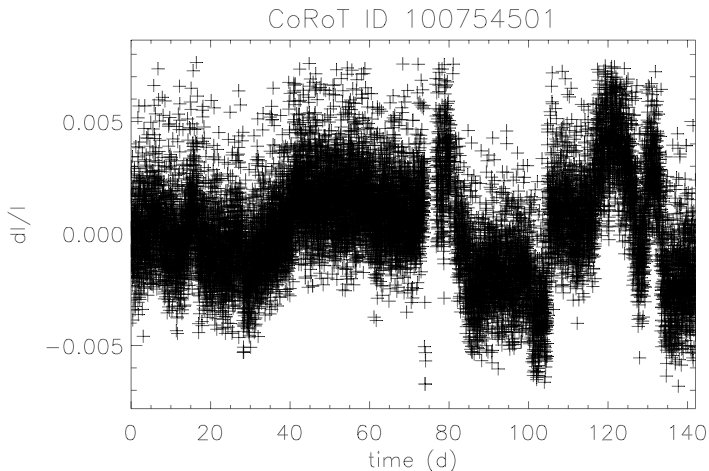}
\includegraphics[angle=0,width=4.5cm]{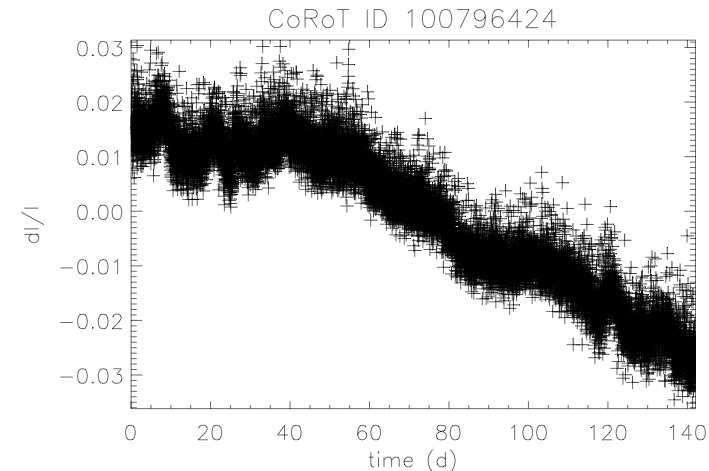}
\includegraphics[angle=0,width=4.5cm]{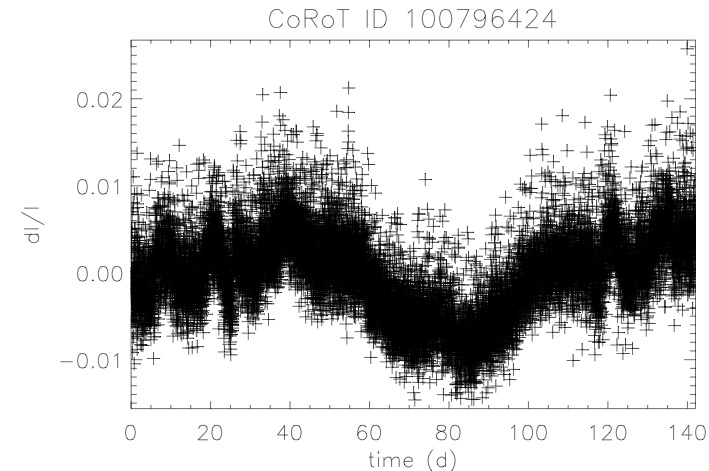}
\includegraphics[angle=0,width=4.5cm]{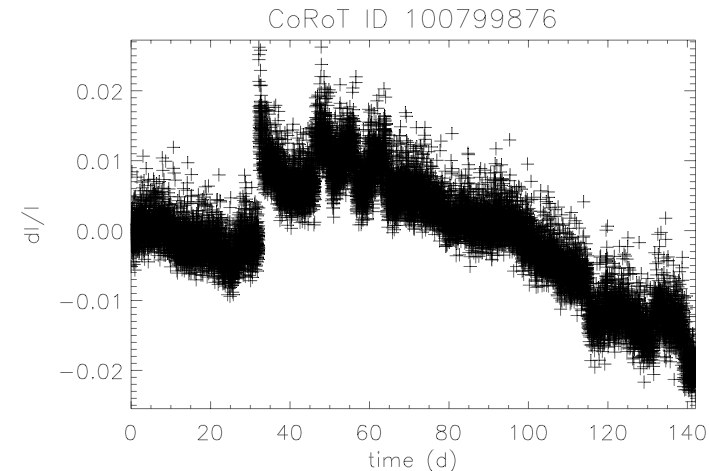}
\includegraphics[angle=0,width=4.5cm]{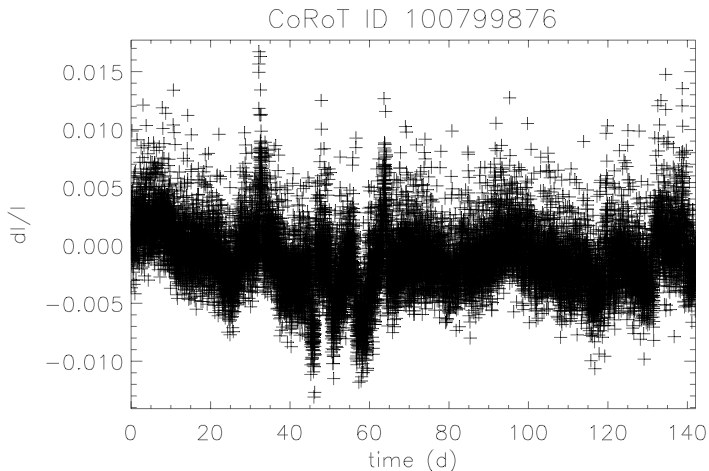}
\includegraphics[angle=0,width=4.5cm]{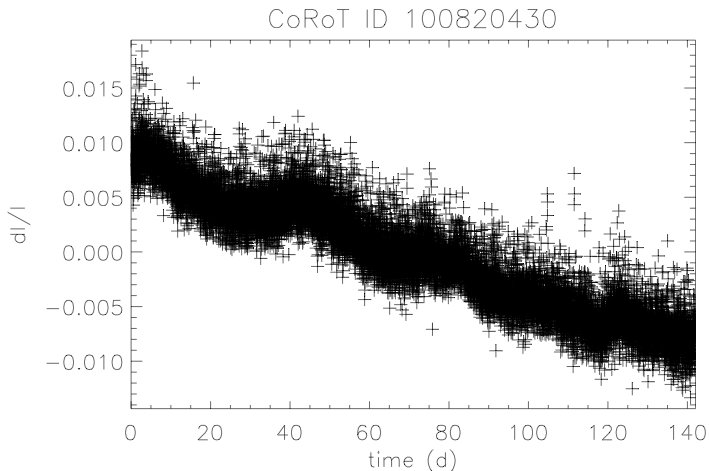}
\includegraphics[angle=0,width=4.5cm]{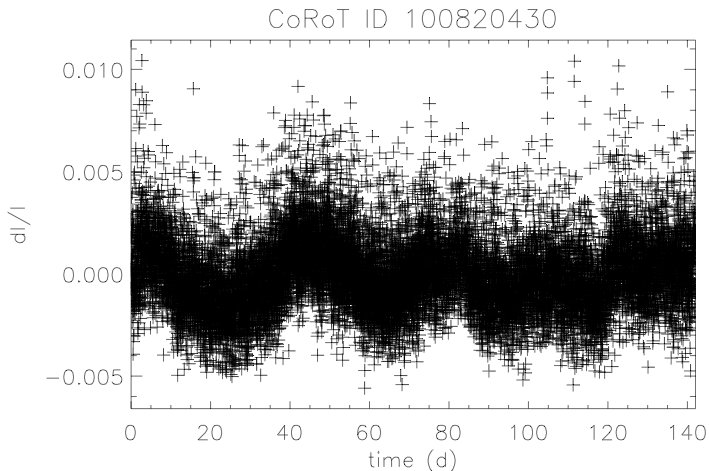}
\includegraphics[angle=0,width=4.5cm]{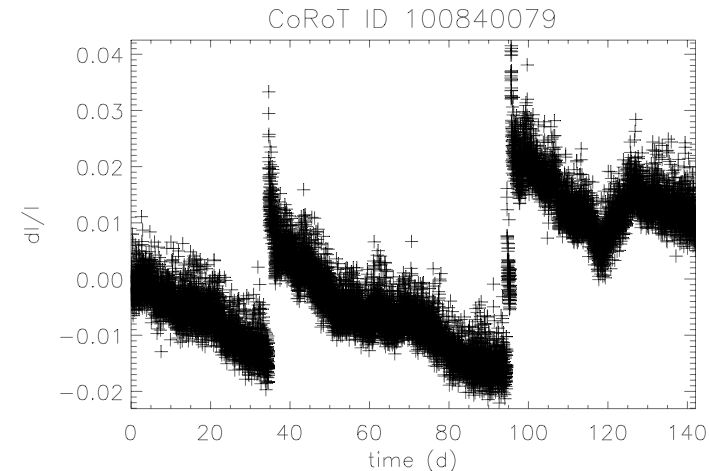}
\includegraphics[angle=0,width=4.5cm]{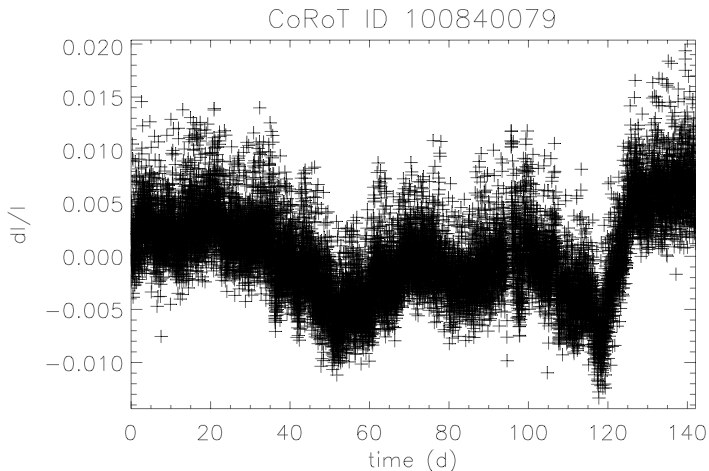}
\includegraphics[angle=0,width=4.5cm]{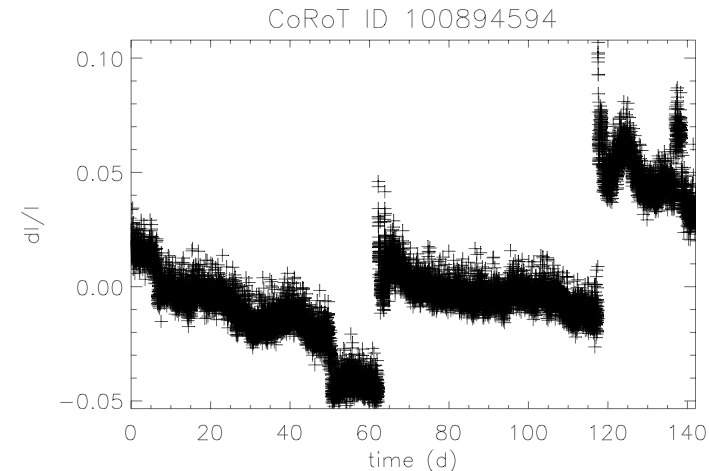}
\includegraphics[angle=0,width=4.5cm]{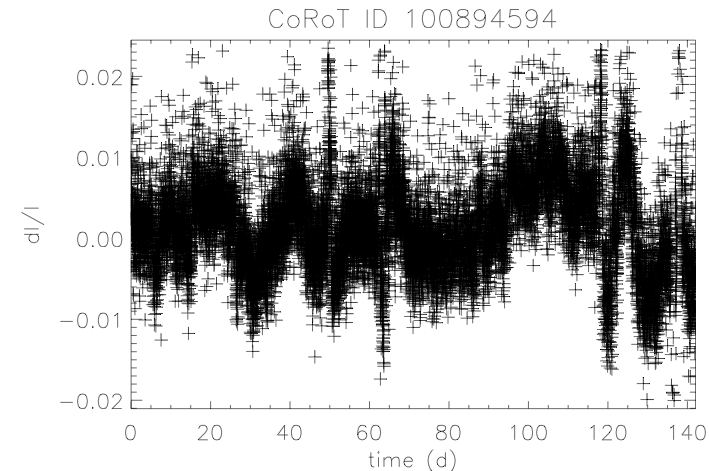}

\caption[]{Plot of  selected light curves  before  and after the removal of the discontinuities 
and/or long-term trends.}
\label{catalogue1}
\end{figure*}

\begin{figure*}[]

\includegraphics[angle=0,width=4.5cm]{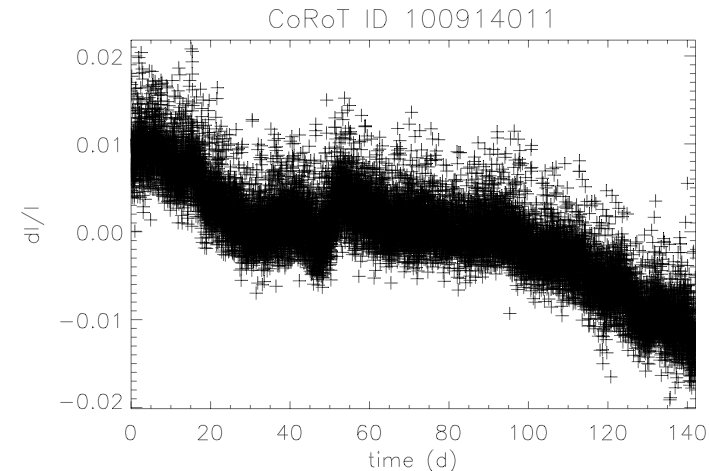}
\includegraphics[angle=0,width=4.5cm]{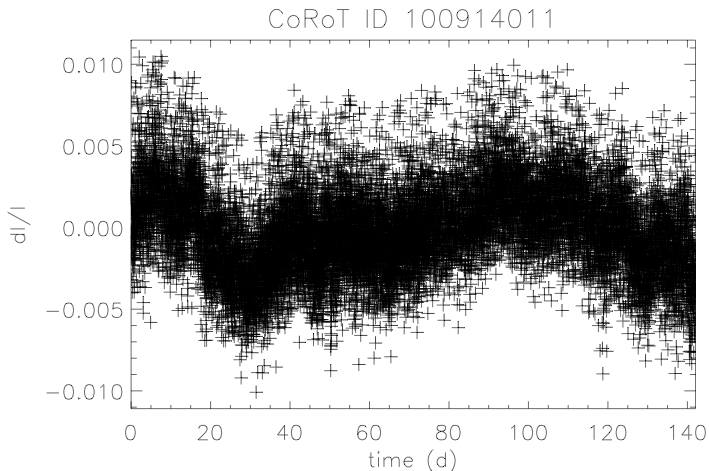}
\includegraphics[angle=0,width=4.5cm]{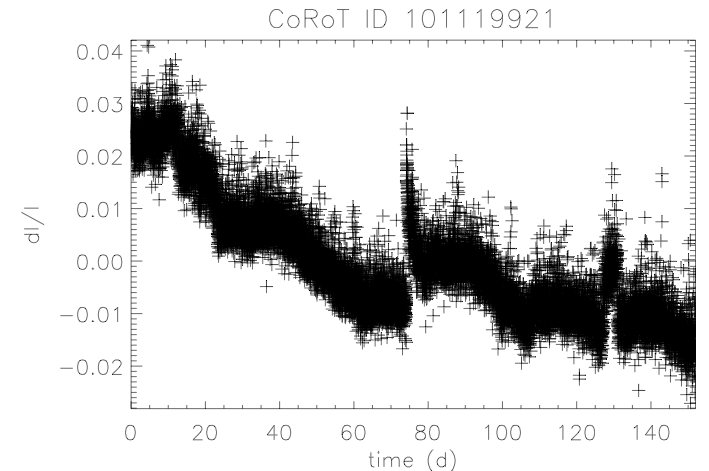}
\includegraphics[angle=0,width=4.5cm]{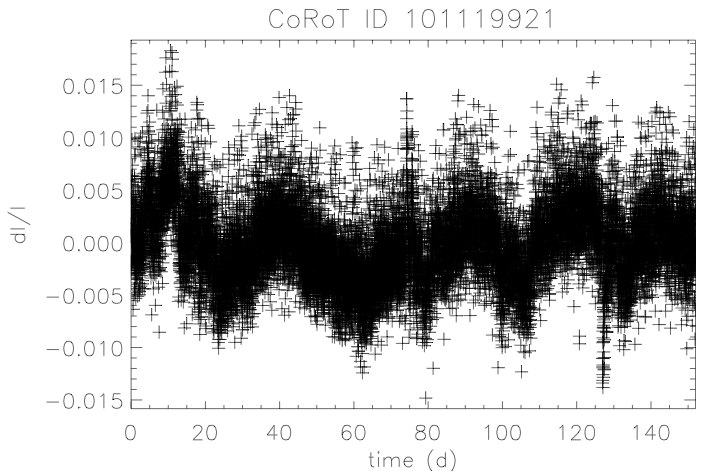}
\includegraphics[angle=0,width=4.5cm]{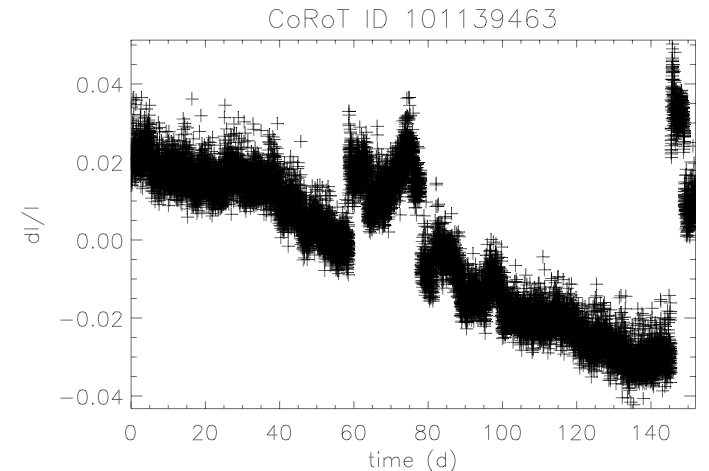}
\includegraphics[angle=0,width=4.5cm]{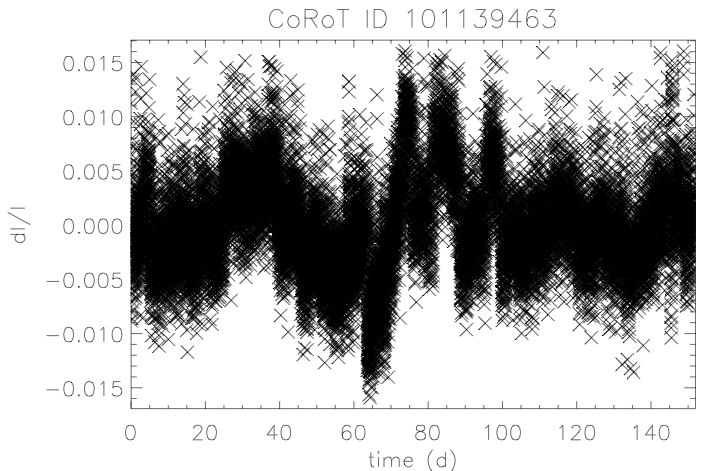}
\includegraphics[angle=0,width=4.5cm]{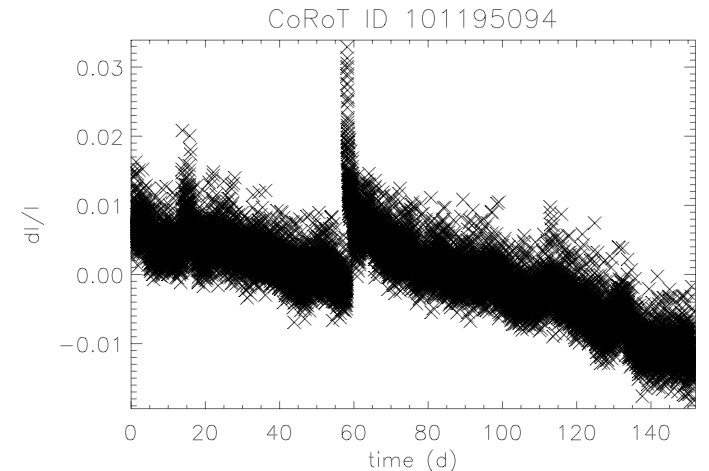}
\includegraphics[angle=0,width=4.5cm]{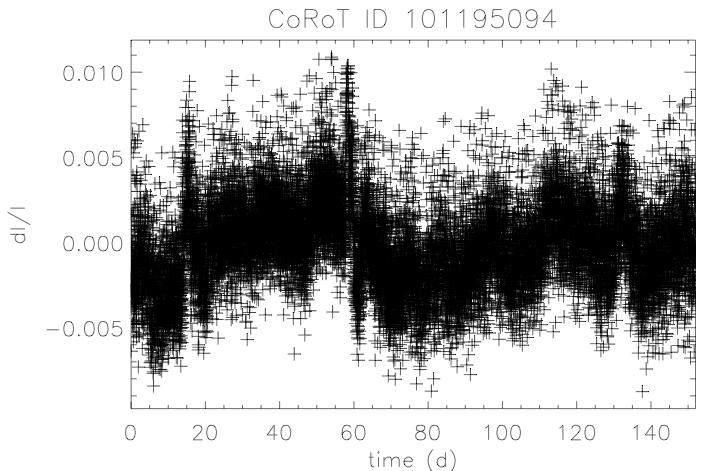}
\includegraphics[angle=0,width=4.5cm]{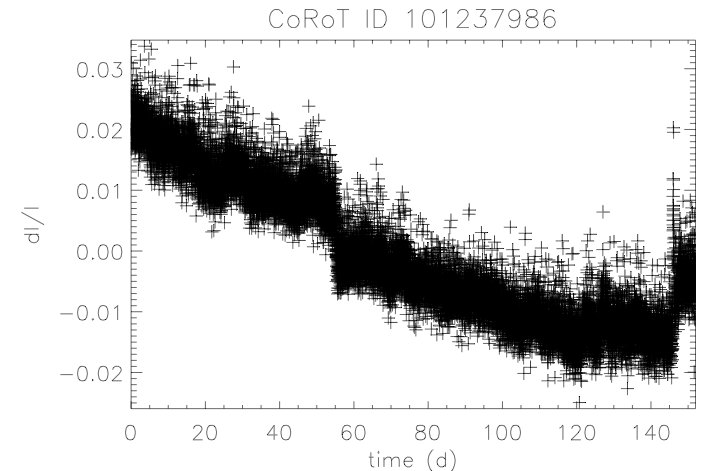}
\includegraphics[angle=0,width=4.5cm]{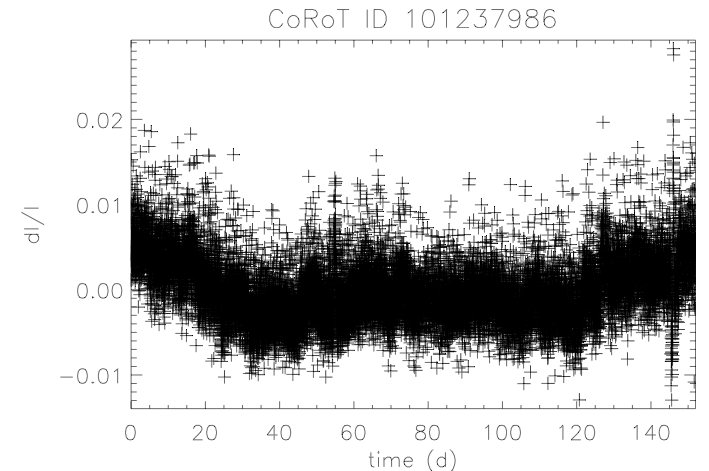}
\includegraphics[angle=0,width=4.5cm]{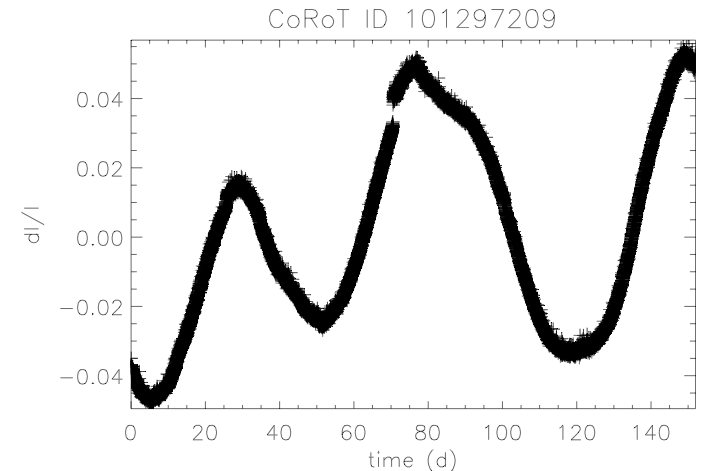}
\includegraphics[angle=0,width=4.5cm]{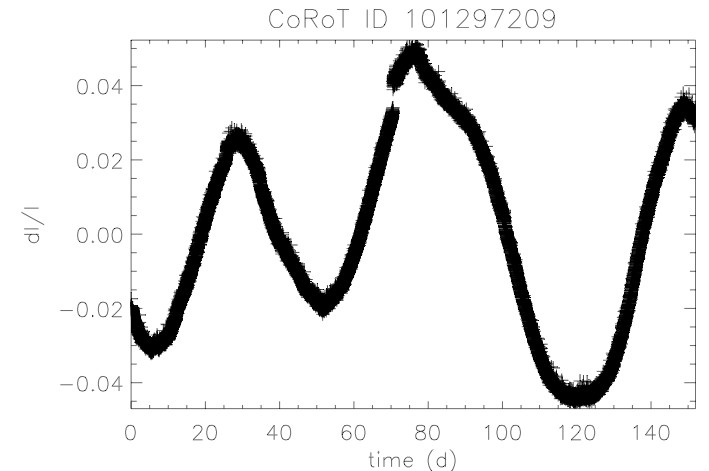}
\includegraphics[angle=0,width=4.5cm]{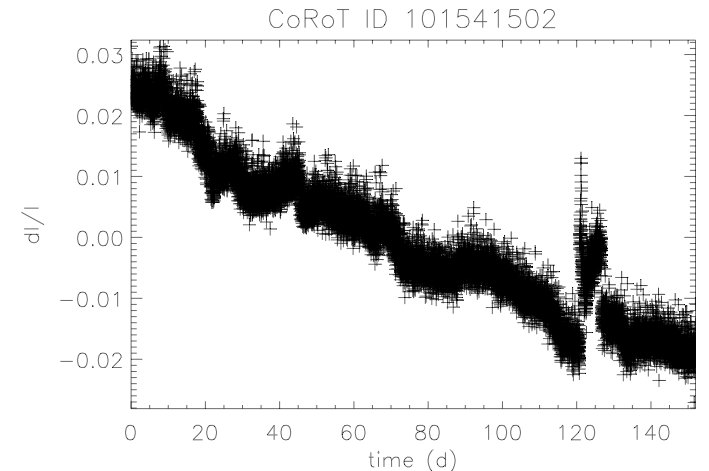}
\includegraphics[angle=0,width=4.5cm]{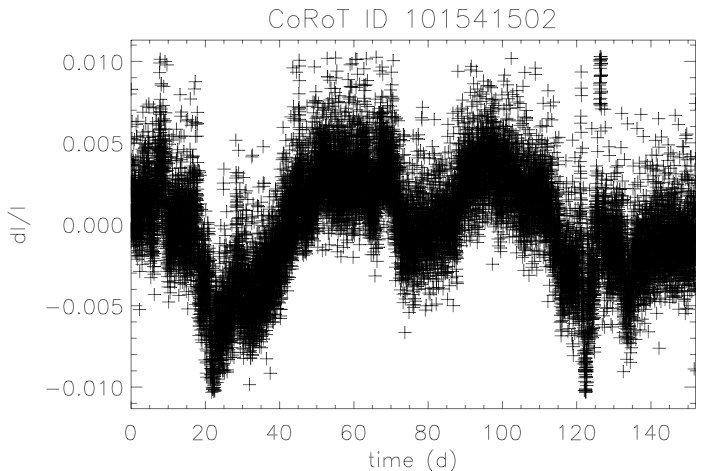}
\includegraphics[angle=0,width=4.5cm]{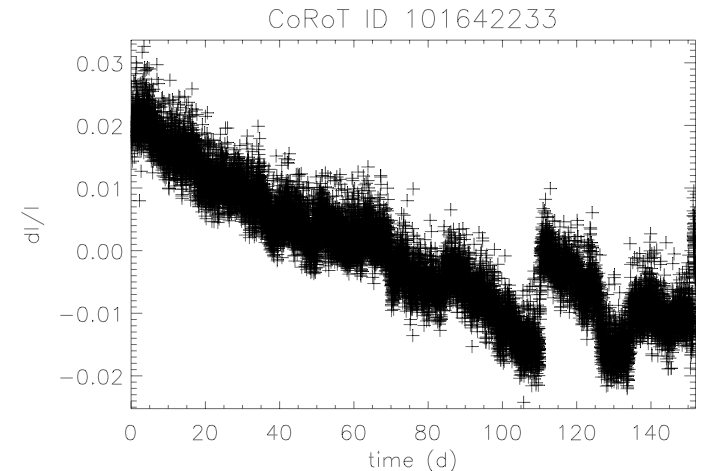}
\includegraphics[angle=0,width=4.5cm]{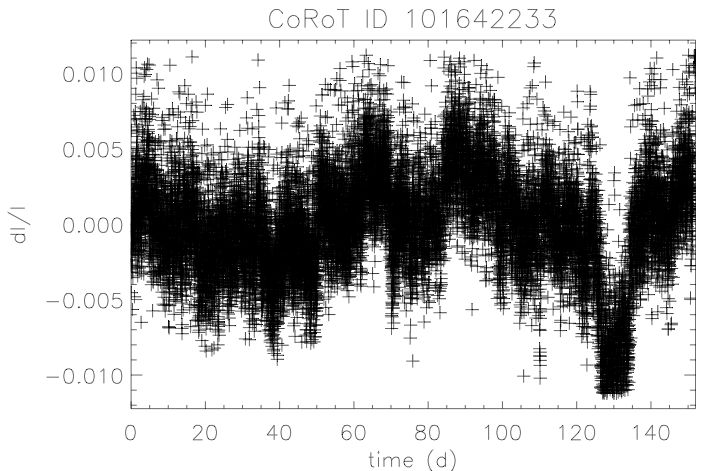}
\includegraphics[angle=0,width=4.5cm]{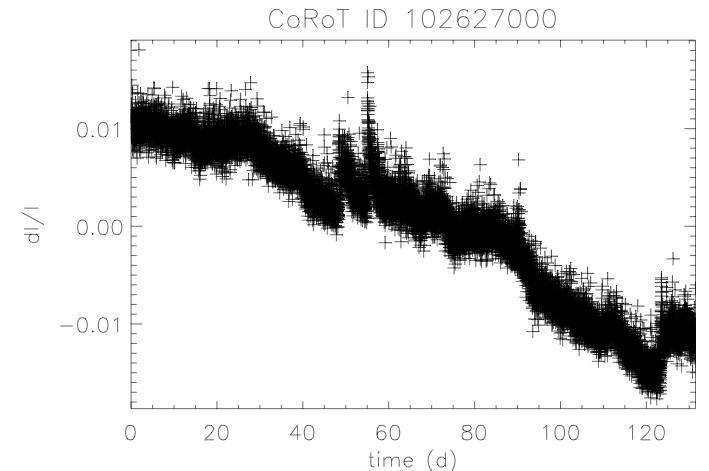}
\includegraphics[angle=0,width=4.5cm]{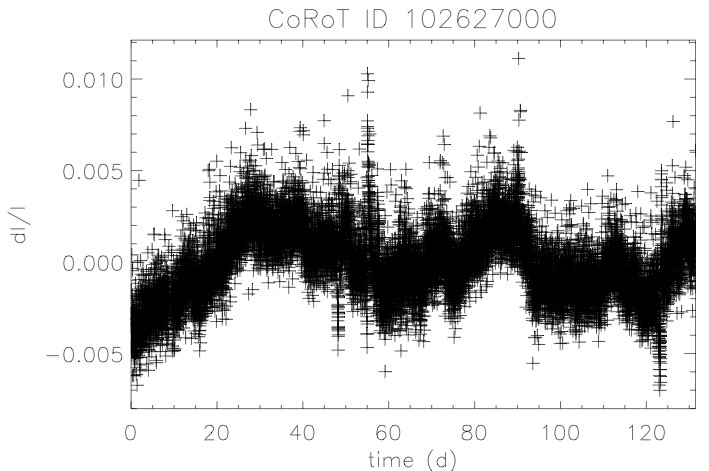}
\includegraphics[angle=0,width=4.5cm]{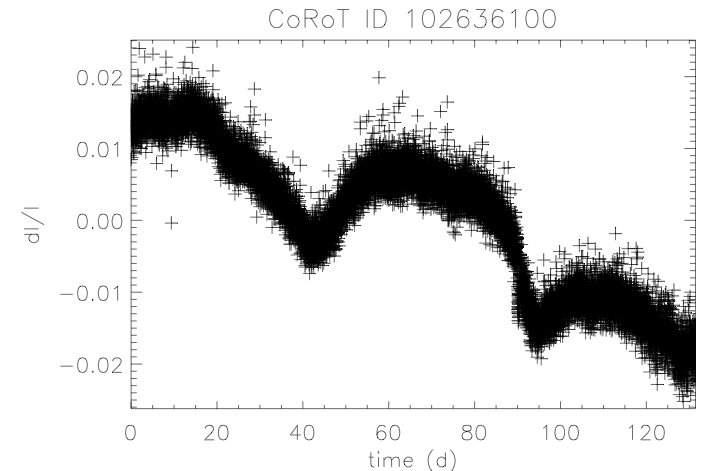}
\includegraphics[angle=0,width=4.5cm]{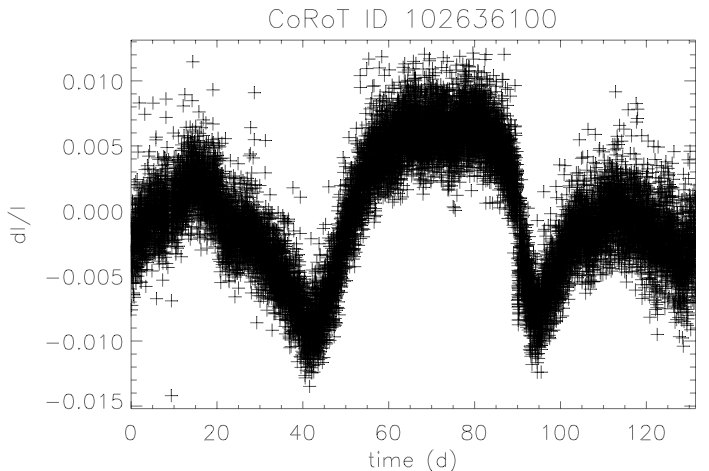}
\includegraphics[angle=0,width=4.5cm]{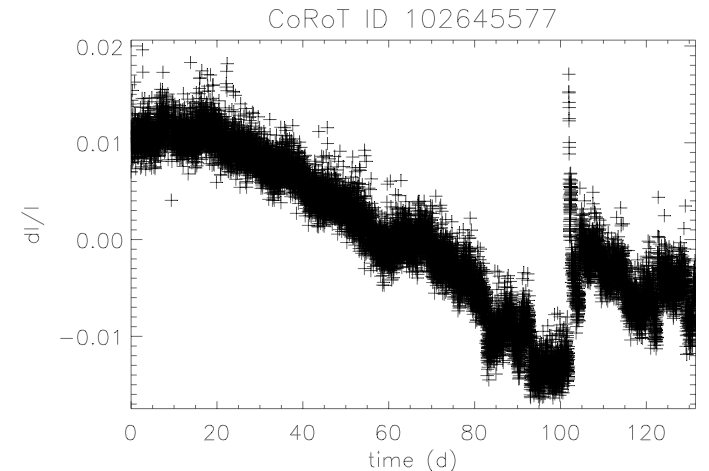}
\includegraphics[angle=0,width=4.5cm]{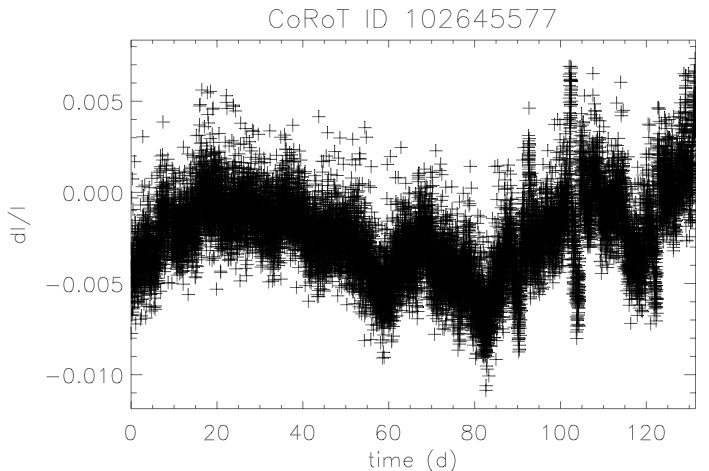}
\includegraphics[angle=0,width=4.5cm]{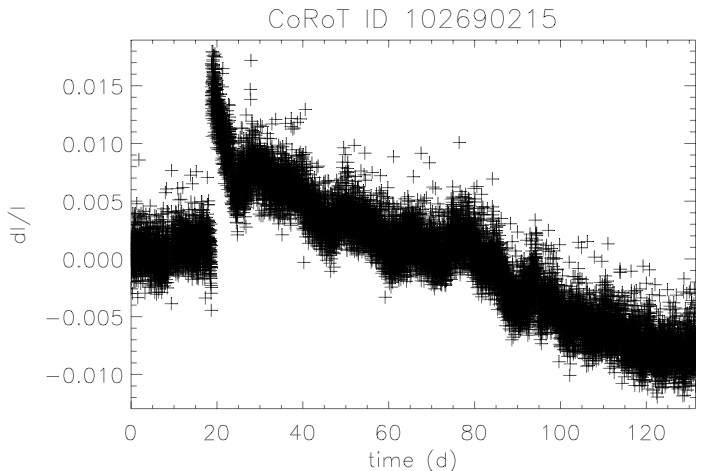}
\includegraphics[angle=0,width=4.5cm]{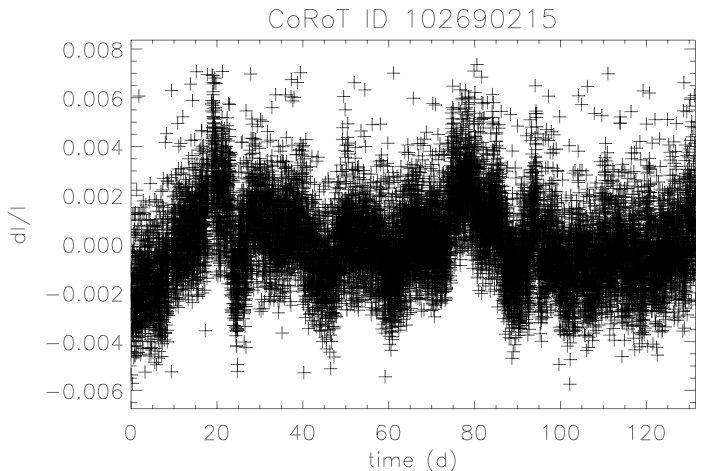}
\includegraphics[angle=0,width=4.5cm]{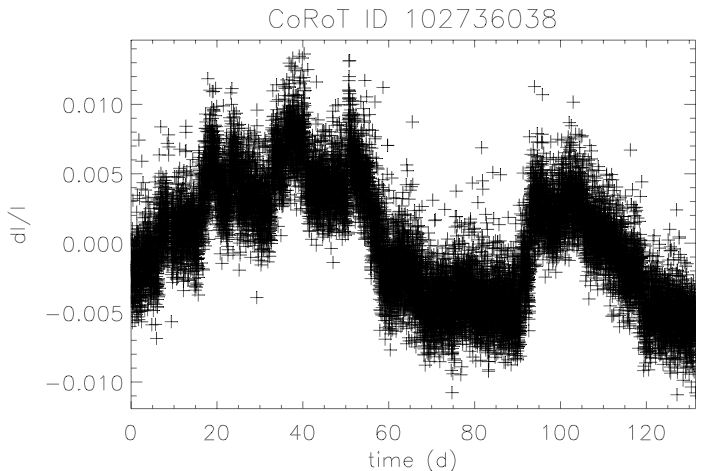}
\includegraphics[angle=0,width=4.5cm]{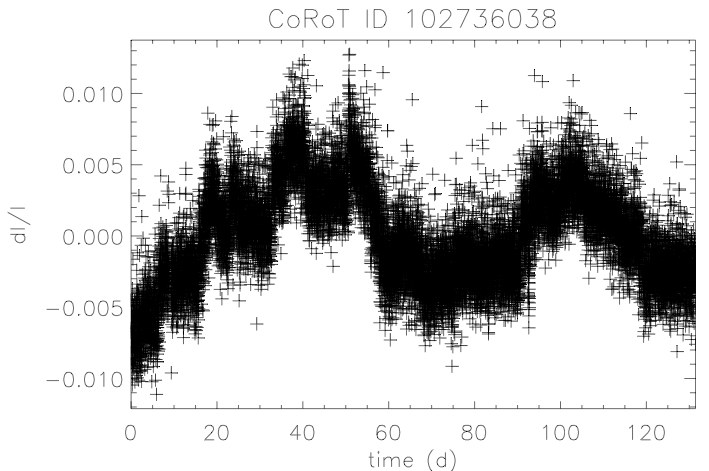}
\includegraphics[angle=0,width=4.5cm]{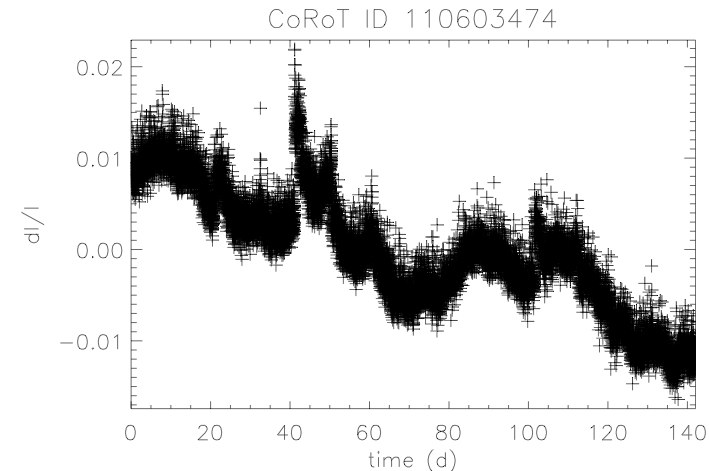}
\includegraphics[angle=0,width=4.5cm]{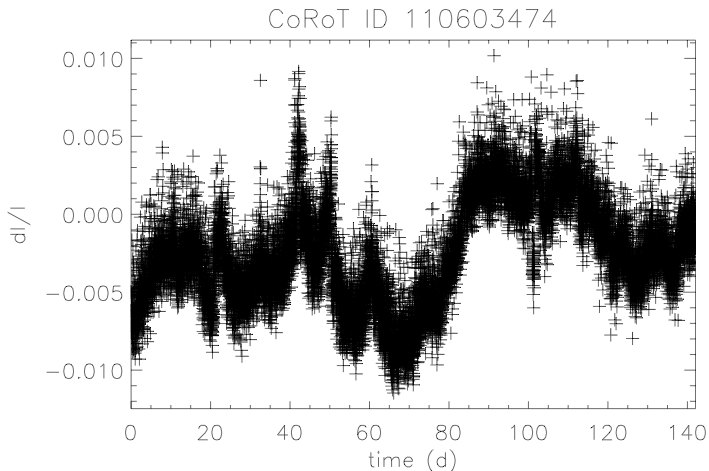}

\caption[]{Plot of  selected light curves  before  and after the removal of the discontinuities 
and/or long-term trends.}
\label{catalogue1}
\end{figure*}

\end{document}